\documentclass[a4paper,UKenglish,cleveref, autoref, thm-restate]{oasics-v2021}
\usepackage{mathtools}
\usepackage{amsmath}
\usepackage{amsthm}
\usepackage{xspace}
\usepackage{listings}
\usepackage{caption}
\usepackage{fancyvrb}
\usepackage{xcolor}
\usepackage{ebproof}
\usepackage{multicol}
\usepackage{enumitem}
\usepackage{arydshln}
\usepackage{booktabs} 

\theoremstyle{definition}
\newtheorem{exmp}{Example}[section]

\newcommand{\toolname}{Netrix\xspace}

\lstdefinelanguage{Tests}{
  keywords={If, Then},
  keywordstyle=\color{red},
  ndkeywords={IsMessageType, MessageTo, And, Or, Gte, MessageFromRound, ReplicaNewRound},
  ndkeywordstyle=\color{blue}\ttfamily,
  emph={DropMessage},
  emphstyle=\color{violet}\ttfamily,
  emph={[2]On},
  emphstyle={[2]\color{brown}\ttfamily},
  emph={[3]Count},
  emphstyle={[3]\color{purple}\ttfamily},
  emph={[4]NewStateMachine, NewTestingServer, NewFilterSet, NewTestCase, AddFilter, SetupFunc},
  emphstyle={[4]\color{olive}\ttfamily},
  identifierstyle=\color{black},
  sensitive=true,
  comment=[l]{//},
  morecomment=[s]{/*}{*/},
  commentstyle=\color{purple}\ttfamily,
  stringstyle=\color{black}\ttfamily,
  morestring=[b]',
  morestring=[b]"
}

\lstdefinelanguage{StateMachine}{
  keywords={NewStateMachine, Builder},
  keywordstyle=\color{red},
  ndkeywords={AddToLog},
  ndkeywordstyle=\color{blue}\ttfamily,
  emph={On, MarkSuccess},
  emphstyle=\color{olive}\ttfamily,
  identifierstyle=\color{black},
  sensitive=true,
  comment=[l]{//},
  morecomment=[s]{/*}{*/},
  commentstyle=\color{purple}\ttfamily,
  stringstyle=\color{black}\ttfamily,
  morestring=[b]',
  morestring=[b]"
}

\lstnewenvironment{syntax}{\lstset{
  basicstyle=\small\ttfamily, 
  language=Tests
}}{}

\DeclareCaptionFormat{listing}{#1#2#3}
\title{A Domain Specific Language for Testing Consensus Implementations}

\author{Cezara Dragoi}{INRIA, ENS}{cezara.dragoi@inria.fr}{}{}

\author{Constantin Enea}{LIX, CNRS, Ecole Polytechnique}{cenea@lix.polytechnique.fr}{}{}

\author{Srinidhi Nagendra}{Universite Paris Cite, CNRS, IRIF, CMI \and \url{http://www.srinidhin.com} }{nagendra@irif.fr}{https://orcid.org/
0000-0002-7171-5543}{}

\author{Mandayam Srivas}{Chennai Mathematical Institute (CMI)}{mksrivas@cmi.ac.in}{}{}

\authorrunning{Dragoi et al.}

\Copyright{Dragoi et al.}



\category{Extended Abstract for Lightning Talk}

\relatedversion{}

\nolinenumbers

\begin{document}

\hideOASIcs
\maketitle

\begin{abstract}
    Large-scale, fault-tolerant, distributed systems are the backbone for many critical software services. Since they must execute correctly in a possibly adversarial environment with arbitrary communication delays and failures, the underlying algorithms are intricate. In particular, achieving consistency and data retention relies on intricate consensus (state machine replication) protocols. Ensuring the reliability of implementations of such protocols remains a significant challenge because of the enormous number of exceptional conditions that may arise in production. We propose a methodology and a tool called Netrix for testing such implementations that aims to exploit programmer's knowledge to improve coverage, enables robust bug reproduction, and can be used in regression testing across different versions of an implementation. As evaluation, we apply our tool to a popular proof of stake blockchain protocol, Tendermint, which relies on a Byzantine consensus algorithm, a benign consensus algorithm, Raft, and BFT-Smart. We were able to identify 4 deviations of the Tendermint implementation from the protocol specification and check their absence on an updated implementation. Additionally, we were able to reproduce 4 previously known bugs in Raft.
\end{abstract}

\section{Introduction}
\label{sec:intro}

Large-scale, fault-tolerant, distributed systems are the backbone of many critical software services. The underlying protocols are intricate as they have to ensure correct behavior in the presence of concurrent asynchronous message exchanges and failures. In particular, \emph{consensus protocols} are used to guarantee data retention and consistency and form the bedrock of storage systems such as Cassandra~\cite{DBLP:journals/sigops/LakshmanM10} and Redis, or Blockchain systems based on proof-of-stake models. Bugs in these implementations have had significant real world consequences.

Ensuring correctness of these implementations remains a significant challenge, precisely because of the enormous number of exceptional conditions that may arise in production, even with a small configuration of hosts and a small number of client requests. Thus, developing effective methodologies for testing distributed systems implementations is an important challenge in systems development.

Testing distributed systems is hard because of the high degree of non-determinism arising out of network and IO scheduling as well as from (benign or Byzantine) faults. An effective testing methodology should provide, at the very least, a reliable way to control the non-determinism---both to enable systematic exploration of the non-deterministic choices and to enable deterministic replay of tests. This by itself is not sufficient. Since the state space is very large, exhaustive or ``blind'' random exploration would typically cover a minuscule fraction of behaviors. Therefore, it is also important to provide the developers an intuitive way to guide the search to interesting \emph{scenarios}: a sequence of ``waypoints'' that describe an interesting behavior of the system.

In this paper, we present \toolname, a domain specific language (DSL) with a networking infrastructure for programmer-guided exploration of large-scale distributed systems implementations. \toolname has two components.

The first is a networking infrastructure to capture and control network events in a deployed system. \toolname runs a central server collecting messages and other events from replicas. It transparently replaces the messaging layer (RPC) of an implementation under test and reroutes the communication primitives to go through the server. 

The second is a domain-specific language (DSL) to write high-level scenario-based unit tests.A unit test in \toolname consists of a set of \emph{filters} and a \emph{state machine} that monitors the execution for some safety or bounded liveness property. A filter imposes restrictions on the order in which \emph{some} messages are delivered by the central server. For example, a filter can specify that a replica is isolated for a selected part of the execution, or that certain protocol messages are dropped, delayed, or modified. Syntactically, filters are written using if-then constructs akin to match-action tables from network devices.

The combination of filters and monitors allow describing semantic scenarios such as creating competition between replicas in a leader election phase by introducing faults. The ability to guide tests to such scenarios is important in testing---in contrast, a pure random exploration may hit such special cases very rarely, if at all. Prior work on testing distributed protocol implementations focuses extensively on \emph{exhaustive enumeration} or \emph{random sampling} of executions for a fixed set of client requests, e.g.,~\cite{DBLP:conf/popl/AbdullaAJS14,modist,samc,conf/cav/YuanYG18,DBLP:journals/pacmpl/OzkanMO19,DBLP:conf/eurosys/LukmanKSSKSPTYL19,DBLP:conf/nsdi/KillianAJV07,DBLP:conf/fast/DeligiannisMTCD16}. Two prominent approaches Jepsen~\cite{jepsen} and PCTCP~\cite{pctcp} have been effective in discovering bugs in production systems. However, the main drawback with the approaches is that they are unreliable in reproducing bugs precisely due to low probabilities of exploring such executions.

While filters constrain the behavior of some messages, the order in which the rest of the messages are delivered is left open. This residual non-determinism is handled by a randomized exploration engine. In our implementation, we use the PCTCP algorithm as the random exploration engine \cite{pctcp}---we show that constraining the sample space of executions using filters leads to a significant improvement in finding bugs in practical systems over unconstrained random exploration.

In summary, filters and monitors provide an effective, semantics-guided, way for programmers to constrain the search space in a test. The random exploration engine complements programmer effort by increasing the range of behaviors specified by the filters. Together, \toolname allows programmers to determine a sweet spot between semantic constraints that determine ``interesting scenarios'' and the search space of message and fault orderings, leading to more effective test plans.

\toolname is implemented in the \texttt{go} programming language. Replicas communicate events and messages over RPC. As a consequence, developers can test implementations written in any programming language by adding a shim over the communication primitives of the implementation. We provide language specific libraries to aid the shim instrumentation effort.

We evaluated \toolname on several production implementations of consensus protocols. Consensus protocols work in the context of a number of servers called replicas that receive requests from clients, and they enable the replicas to agree on a common order in which to execute the client requests. During the execution of the protocol, replicas maintain a state and change it when they send/receive messages. The protocol defines rules to modify this state based on the type, order, and number of messages that arrive at a replica. Links between processes may fail, messages can be lost or delayed, and protocols must ensure consensus in the presence of such failures.

We used \toolname to test several large-scale implementations: an open-source implementation of the Raft~\cite{raft} protocol,\footnote{\url{https://github.com/etcd-io/etcd/tree/main/raft}} the Tendermint~\cite{tendermint} protocol,\footnote{\url{https://github.com/tendermint/tendermint} deployed in production in the Cosmos ecosystem that has over a hundred blockchains} and BFT-Smart~\cite{DBLP:conf/dsn/BessaniSA14,DBLP:conf/edcc/SousaB12}.\footnote{\url{https://github.com/bft-smart/library}}
 Using \toolname, we were able to (1) write unit tests to explore behaviors that are not commonly observed in production environments, including known bugs; (2) demonstrate behaviors where the implementation deviates from the protocol specification; and (3) run tests on multiple versions of the implementation, thereby checking corrections for the observed deviations. 
In particular, our tests for Tendermint allowed the Tendermint developers to gain confidence in the correctness of the implementation and write further unit tests for planned changes. 

Overall, we make the following contributions,
\begin{itemize}
    \item We introduce a DSL (filters and monitors) to write reusable and reproducible unit tests for testing distributed protocol implementations. 
    \item We provide an infrastructure for executing the unit tests by controlling the communication layer.
    We demonstrate through a number of experiments that the combination of our DSL and randomized exploration can provide effective testing of large-scale systems.
    \item We evaluate our tool and DSL on large-scale, deployed, implementations of \textbf{Raft}, \textbf{Tendermint}, and \textbf{BFT-Smart}. In particular, we report our findings for Tendermint, where we identified 4 bugs and helped the developer team to check fixes to the bugs.
\end{itemize}

\paragraph{Outline} We provide an overview of \toolname unit tests in Section~\ref{our_approach}. Section~\ref{system_model} provides the theoretical semantics of an execution under \toolname, while Section~\ref{syntax} explains the syntax and semantics of our DSL. We present our findings and common testing patterns while testing Tendermint, Raft and BFTSmart in Section~\ref{case_study}. In Section~\ref{related_work}, we present other related work before concluding in Section ~~\ref{conclusion}.
\section{Our approach}
\label{our_approach}

\begin{figure}[t]
  \centering
  \includegraphics[width=0.6\linewidth]{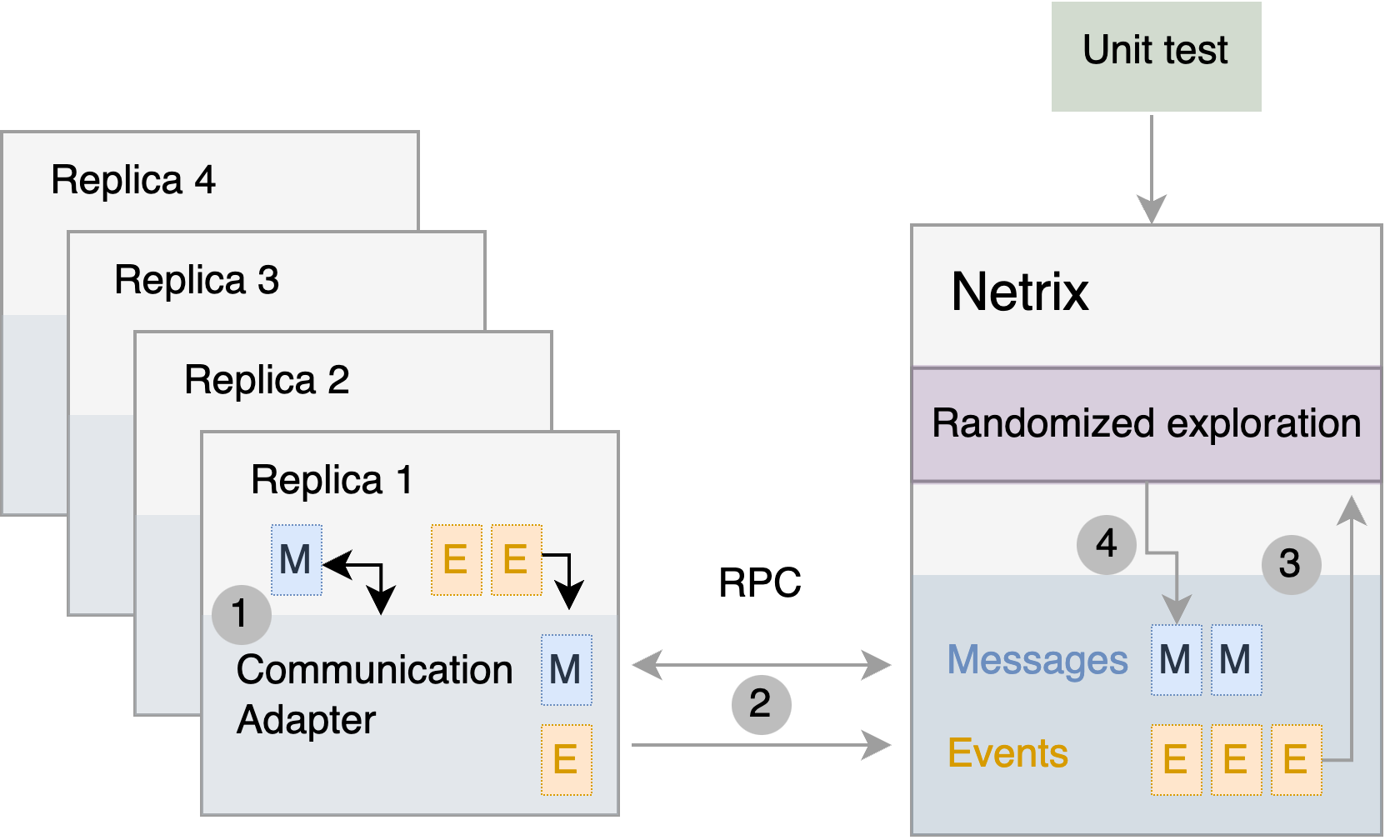}
  \caption{\textbf{\toolname architecture}}{(1) The replicas submit events, send/receive messages to a communication adapter. (2) The adapter talks to \toolname via RPC. (3) The events and messages drive unit tests. (4) Unit tests decide which messages to deliver}
  \label{fig:overview_arch}
\end{figure}

We test with \toolname distributed systems where processes communicate via message passing. An execution in such a system is characterized by a sequence of events at each process. An event may be either send/receive a message or other internal steps such as start/end a timer. Figure~\ref{fig:overview_arch} illustrates the architecture of \toolname applied to a 4-node distributed system. \toolname tests directly the system implementation, using a communication adapter for controlling the network. 

\toolname determines the order in which messages are received using two mechanisms. The first and default mechanism is a randomized exploration algorithm (in our case PCTCP) for sampling executions of a distributed system. Second, \toolname receives as input a unit test written by a developer. The unit test processes the events as a stream and determines the messages to be delivered at each step. A unit test can, for instance, block delivery of a message based on event A and deliver the message upon observing a later event B. These blocked messages are not seen by the random exploration algorithm. Additionally, a unit test can also introduce fictitious messages which simulate Byzantine behaviors (e.g., where replicas send messages that are not prescribed by the protocol). A unit test consists of a set of filters to control message delivery and an assertion defined by a state machine. We introduce a domain-specific language that is embedded in \texttt{go} to specify filters and assertions. Filters allow developers to introduce specific faults such as dropping messages, reordering, or replaying messages. If no filters are specified, the delivery of messages is controlled entirely by the random exploration algorithm.

\begin{figure}
  \centering
  \includegraphics[width=0.5\linewidth]{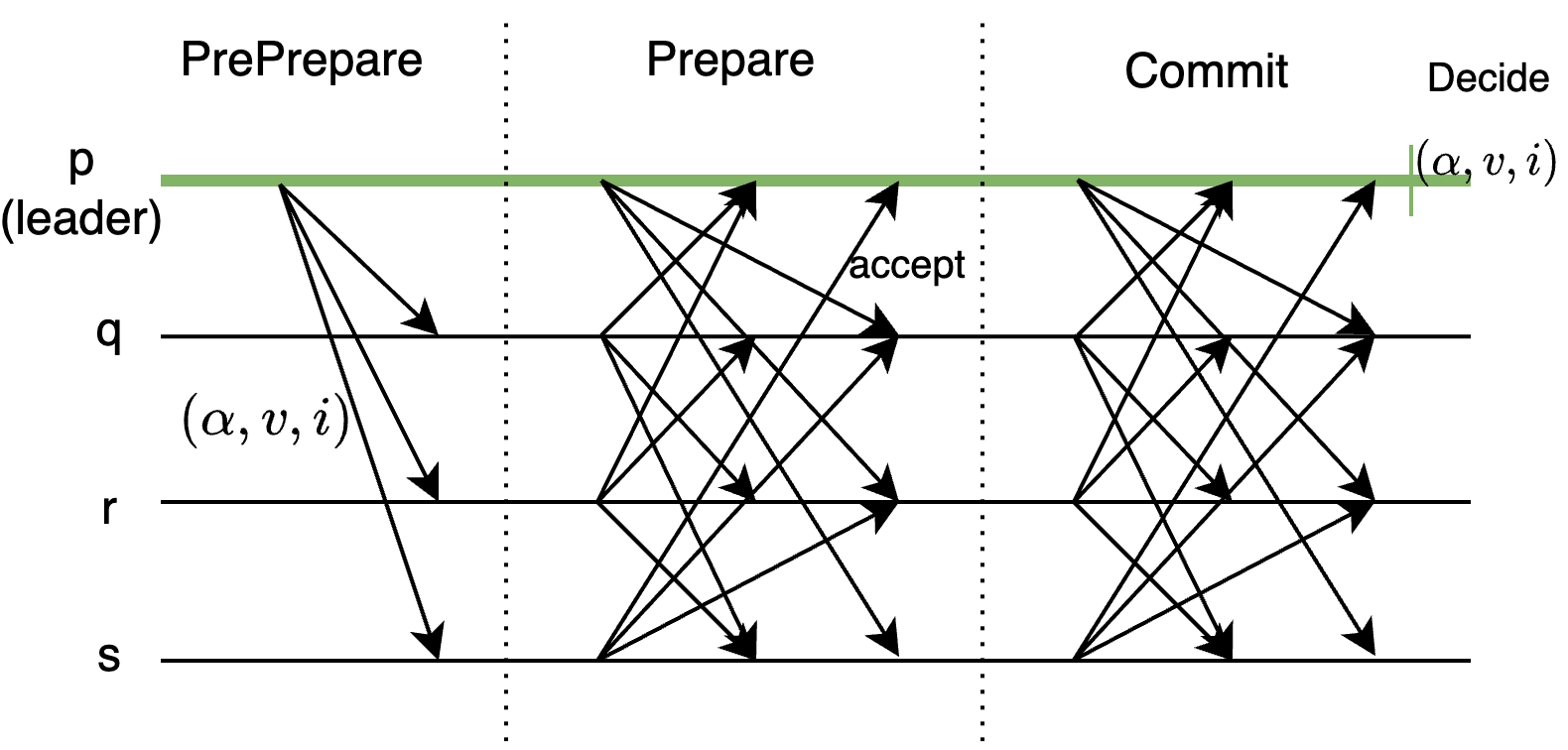}
  \caption{PBFT normal case execution with $p$ as leader.}
  \label{fig:pbft_normal}
\end{figure}

\begin{figure}
  \centering
  \includegraphics[width=0.5\linewidth]{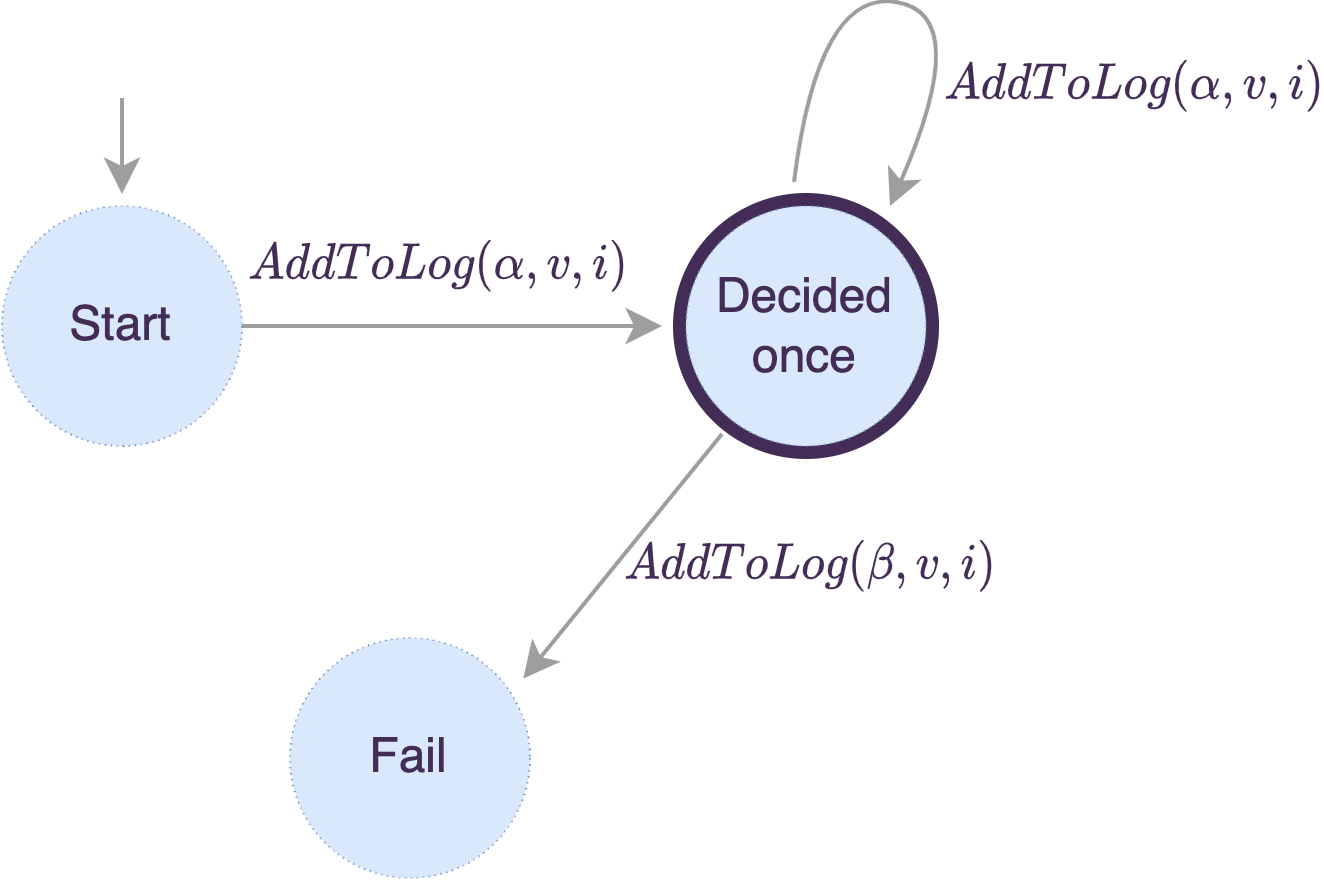}
  \caption{Event driven state machine for assertions.}
  \label{fig:pbft_property_sm}
\end{figure}

To explain the structure of filters and assertions, we use the Practical Byzantine Fault Tolerance (PBFT)\cite{pBFT} protocol as a running example for the rest of the paper. PBFT solves consensus in the presence of Byzantine faults. For a distributed system of $n$ replicas, PBFT can tolerate $f < \frac{n}{3}$  byzantine replicas. Replicas go through a sequence of rounds of communication called views. In each view $v$, one replica is chosen to be the leader, and the remaining replicas are identified as backups.

Figure~\ref{fig:pbft_normal}, illustrates the ``normal'' execution of the protocol (no faults and instantaneous message delivery). In the \textit{PrePrepare} phase, the leader $p$ accepts a request $\alpha$ from the client and broadcasts a  \texttt{PrePrepare($\alpha,v,i$)} message to all replicas. By doing so, the leader proposes to add the request $\alpha$ at index $i$ to the log in view $v$. In the \textit{Prepare} phase, if a replica accepts the \texttt{PrePrepare} message, it broadcasts a \texttt{Prepare($\alpha,v,i$)} message. Once a replica receives $2f+1$ \texttt{Prepare} messages matching the \texttt{PrePrepare} message, it proceeds to the \textit{Commit} phase and broadcasts a \texttt{Commit($\alpha,v,i$)} message. Finally, upon receiving $2f+1$ \texttt{Commit} messages that match the \texttt{Prepare} message, each replica decides to add the request to the log.  The client sends the request $\alpha$ to all replicas and upon receiving the request, the replicas start a timer. When the timer expires, if the replica has not added $\alpha$ to the log a view change protocol is initiated by broadcasting a \texttt{ViewChange($v+1$)} message for the next view. When the leader of the view $v+1$ receives \texttt{ViewChange} messages from $2f$ other replicas, it broadcasts \texttt{NewView($v+1$)} messages to all replicas.

The correctness of PBFT relies on the property that no two replicas decide to add different requests to their log for a given index $i$ in any given view. Developers can write state machines to specify this property. Consider a system of 4 replicas that receives two client requests $\alpha$ and $\beta$. We should not observe a replica $p$ adding $\alpha$ at index $i$ and replica $q$ adding $\beta$ at the same index. Figure~\ref{fig:pbft_property_sm} illustrates the state machine for this property. The transitions between states are labelled by a condition on an event. Starting from the initial state, 
the state machine takes a step every time we process an event received from a replica. When a transition condition is satisfied for the current event, we transition to the next state. In this case, \texttt{AddToLog($\alpha$,v,i)} is a condition that is satisfied on the event indicating a replicas decision to add a request ''\texttt{$\alpha$}'' to the log at index ''\texttt{i}'' in view ''\texttt{v}''. The predicate can be a function on the messages that~\toolname observes, or the implementation explicitly emits the event (part of the instrumentation effort). The first time we observe such an event for a client request ''\texttt{$\alpha$}'' we transition to an accepting ''Decided Once'' state. If we observe an event indicating a decision for a different client request ''\texttt{$\beta$}'' then we transition to a fail state.

Unit tests that contain no filters capture executions with arbitrary communication delays and faults, but no Byzantine faults. All messages that are delivered were sent by replicas following the protocol and the order in which messages are delivered is under the control of the underlying random exploration algorithm. Filters can be used to introduce specific delays or faults whose probability of being exposed by a random exploration algorithm is very low, or fictitious messages that simulate Byzantine faults. For instance, we can drop \texttt{Prepare} messages to one replica $p$ and check if the property stated above still holds. Example~\ref{exmp:benign_drop_1} describes the corresponding filter.

\begin{figure}
  \begin{exmp}
  \label{exmp:benign_drop_1}
  Drop \texttt{Prepare} messages of view $v$ to one specific replica $p$
  \begin{verbatim}
    If(IsMessageOfView(v)
      .And(IsMessageType(Prepare))
      .And(MessageTo(p))
    ).Then(DropMessage())
  \end{verbatim}
  \end{exmp}
\end{figure}

\begin{figure}
  \begin{exmp}
  \label{exmp:benign_drop_2}
  Drop all \texttt{Prepare} messages of view $v$
  \begin{verbatim}
    If(IsMessageOfView(v)
        .And(IsMessageType(Prepare)
    ).Then(DropMessage())
  \end{verbatim}
  \end{exmp}
\end{figure}

Filters form a general structure of, \texttt{If(condition).Then(actions)}. In this case, we define the conditions \texttt{IsMessageType} and \texttt{MessageTo} and compose them with logical operators. The condition \texttt{IsMessageType} is true when the event is of type message-send, and the message is of the specified type. \texttt{DropMessage} is an action that does not deliver the message and consequently, will not be seen by the random exploration algorithm. When the condition is true, the corresponding action is executed. Here, we drop \texttt{Prepare} messages to $p$. Consequently, $p$ will request for view change as it does not add the request $\alpha$ to its log. We expect the leader of the new view to not initiate a view change as only one out of the four replicas send a view change request. 

\begin{figure}
    \centering
    \begin{subfigure}{0.48\linewidth}
        \includegraphics[width=\textwidth]{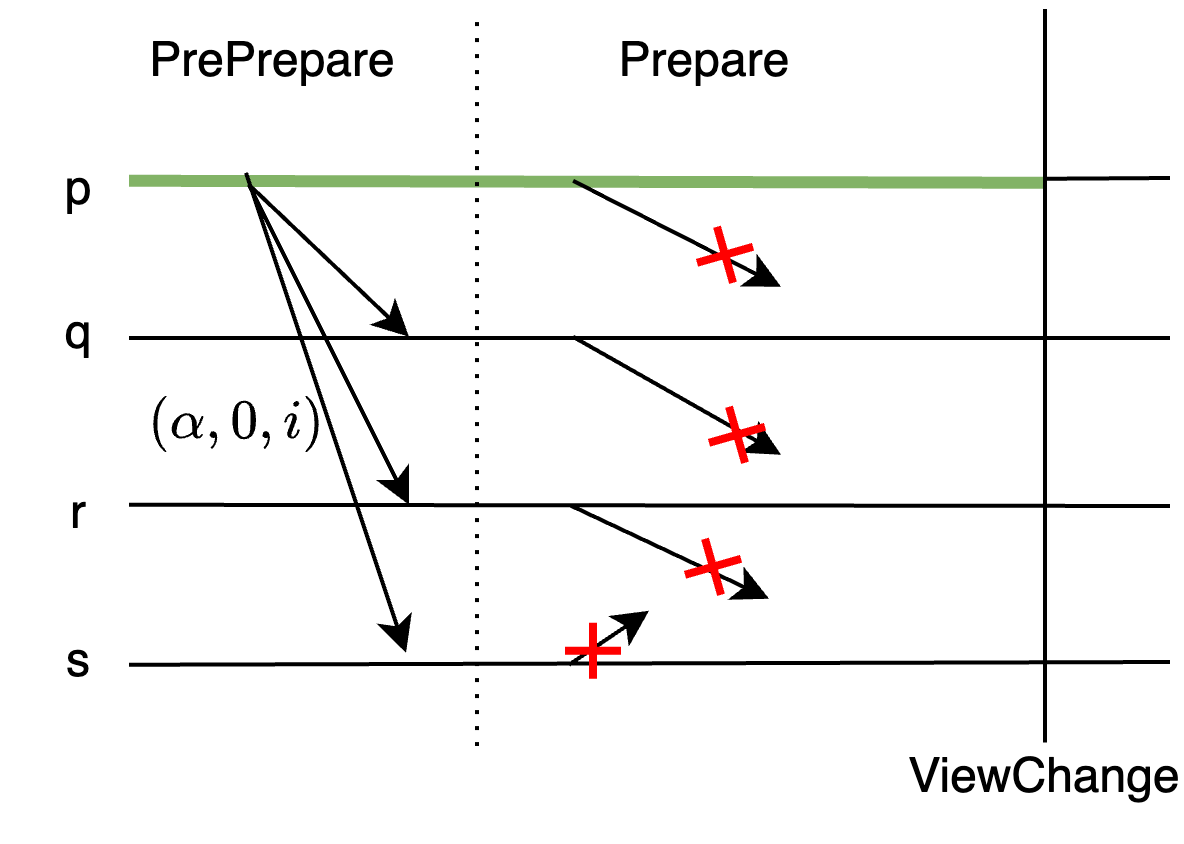}
        \caption{Dropping all \texttt{Prepare} messages}
    \end{subfigure}
    \hfill
    \begin{subfigure}{0.48\linewidth}
        \includegraphics[width=\textwidth]{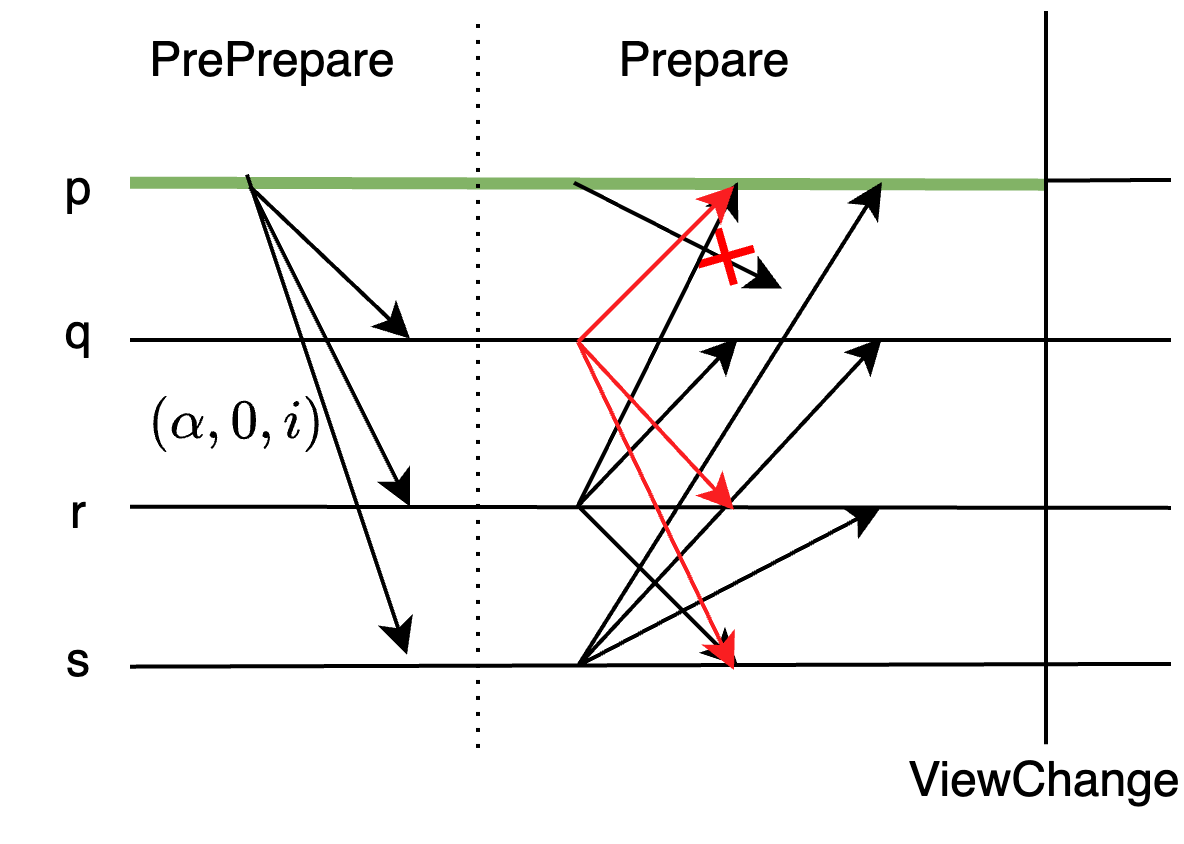}
        \caption{Introducing byzantine faults by change \texttt{Prepare} to nil}
    \end{subfigure}
    \caption{Two different scenarios where we use filters to force a view change}
    \label{fig:pbft_vc}
\end{figure}

Alternatively, developers can enforce certain execution paths using filters. For example, let us look at different ways to force a view change. One, by dropping all \texttt{Prepare} messages using just one filter (Example~\ref{exmp:benign_drop_2}). Two, by dropping \texttt{Prepare} messages of one replica $p$ and change all \texttt{Prepare} messages from $q$ to $nil$ (Example ~\ref{exmp:byzantine}). This demonstrates the filters ability to introduce byzantine behaviour. In both the scenarios replicas will not receive sufficient \texttt{Prepare} messages to make progress and commit and hence initiate a view change. Note that in Example~\ref{exmp:byzantine}, we use a custom action \texttt{ChangePrepareToNil}. Filters can use custom actions that are specific to the protocol implementation. Figure~\ref{fig:pbft_vc} illustrates the two scenarios where we force a view change.

The state machine for Example~\ref{exmp:benign_drop_2} and~\ref{exmp:byzantine} is strengthened to check if additionally the replicas initiated a view change. This is an instance of a (bounded) liveness property which requires that something ``good'' should eventually happen (it is bounded liveness because it is checked on executions that finish before a timeout expires -- see Section~\ref{syntax}). More precisely, from the initial state, upon observing 3 \texttt{ViewChange(1)} messages from distinct replicas, we transition to "ViewChangeExpected" state. Subsequently, when we observe a \texttt{NewView(1)} message from the leader of view 1, we transition to an accepting state.

A random exploration algorithm is very unlikely to sample executions where, like in Example~\ref{exmp:benign_drop_2}, messages are dropped or delivered towards the end. The low probability of delivering these messages at the end is due to the large number of messages in an execution (in the order of hundreds). Filters provide developers with a mechanism to introduce specific faults, delays and Byzantine behavior. Intuitively, this is analogous to mechanized proofs requiring additional input from the user, e.g., inductive invariants.

\begin{exmp}
\label{exmp:byzantine}
Change \texttt{Prepare} messages of $p$ to $nil$
\begin{verbatim}
  If(IsMessageType(Prepare).
      And(MessageFrom(p))
  ).Then(DropMessage)
  If(IsMessageType(Prepare).
      And(MessageFrom(q))
  ).Then(ChangePrepareToNil())
\end{verbatim}
\end{exmp}

\begin{figure}
  \centering
  \includegraphics[width=0.5\linewidth]{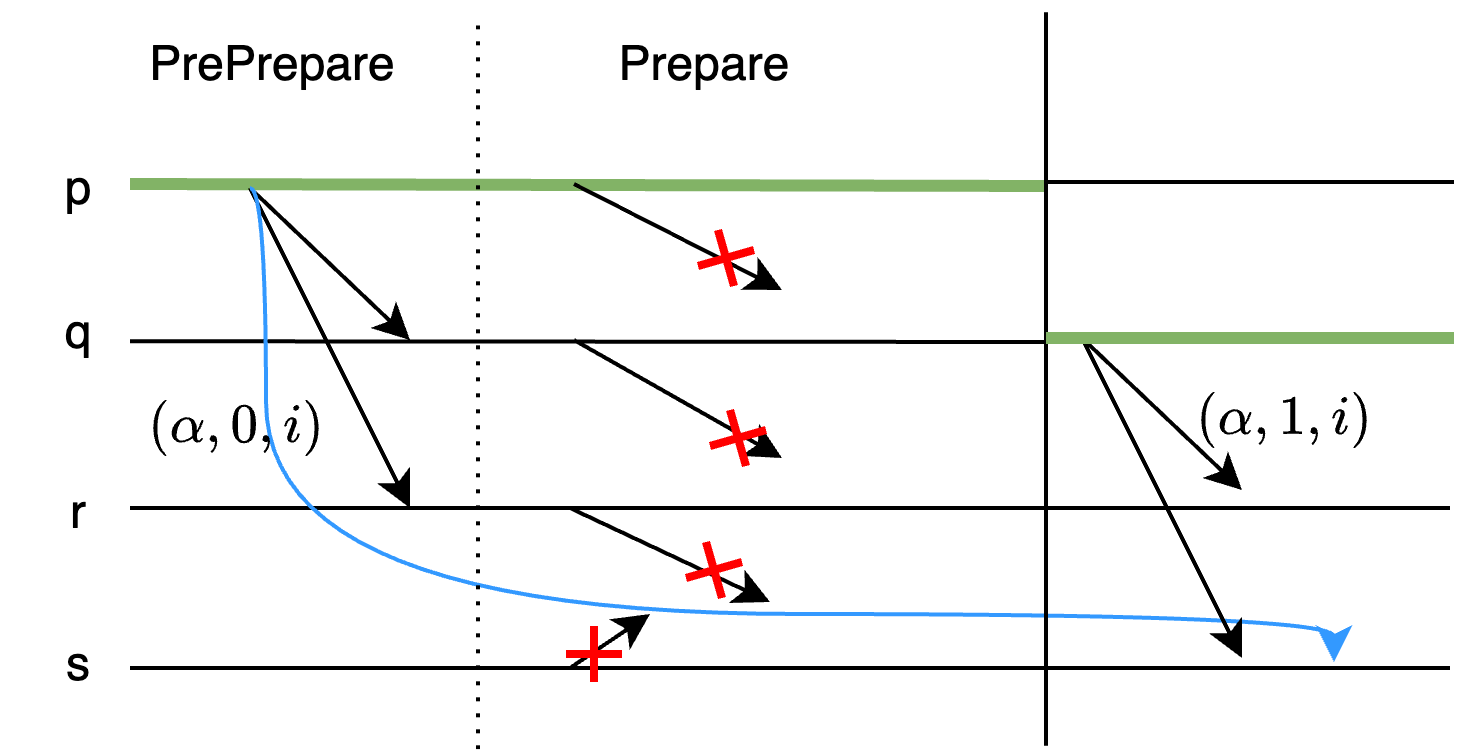}
  \caption{Reordering \texttt{PrePrepare} for $r$  from view 0 until new view.}
  \label{fig:pbft_reorder}
\end{figure}

In addition to introducing Byzantine behavior, the DSL allows reordering messages. Example~\ref{exmp:reorder} describes three filters. (1) The filter drops all \texttt{Prepare} messages of view 0. (2) \texttt{PrePrepare} message of view 0 to a replica $r$ are stored in the "reordered" set. (3) Messages stored in the "reordered" set are delivered when a \texttt{PrePrepare} message of view 1 is observed. Under this scenario, we test for the generic safety property that no two processes decide different values (in the same view). Figure~\ref{fig:pbft_reorder} illustrates the corresponding execution. In this unit test, Filter 1 ensures that processes move to the next view and filters 2 and 3 ensure that process $r$ receives two different \texttt{PrePrepare} messages.

\begin{exmp}
\label{exmp:reorder}
Reorder the \texttt{PrePrepare} message to process $r$ to the next view.
\begin{verbatim}
1. If(IsMessageType(Prepare).And(MessageView(0))).
      Then(DropMessage())
2. If(IsMessageType(PrePrepare).And(MessageTo(r)).And(MessageView(0))).
      Then(StoreInSet("reordered"))
3. If(IsMessageType(PrePrepare).And(MessageView(1))).
      Then(DeliverAllFromSet("reordered"))
\end{verbatim}
\end{exmp}

As mentioned above, a random exploration algorithm is unlikely to explore executions that correspond to Filter 1 (dropping messages), but may be effective in exploring more restricted re-orderings like the one represented by Filters 2 and 3. We observe this in our evaluation as well. Using PCTCP as the underlying exploration algorithm, we see that after running 200 iterations without any of the filters we do not observe any execution where \texttt{Prepare} messages of view 0 are delivered at the end. On the other hand, Filters 2 and 3 define a message re-ordering with a smaller scope: \texttt{PrePrepare} messages of view 0 to a process $r$ need to be delivered after \emph{any} message from view 1 and PCTCP is able to explore this reordering.

This provides a tradeoff to the developers writing unit tests where more expressive tests (in terms of number of filters) provide better probabilistic guarantees. On the other hand, broader exploration of the execution space can be achieved with smaller tests. Furthermore, in Section~\ref{case_study}, we associate filters with a distance score based on the number of messages between the actual position in a synchronous execution and the expected position after re-ordering. We show experimentally that filters that exceed a certain distance threshold are necessary to constrain the execution space to observe the desired property. The distance measure can help developers in making the right tradeoff decisions.

We present the syntax and the semantics of the filters and of the state machine in Section~\ref{syntax} (along with the instrumentation effort). The next section formalizes the semantics of a protocol with \toolname.
\section{System model}
\label{system_model}

% Common variables
\newcommand{\allMessages}{\mathcal{M}}
\newcommand{\allReplicas}{\mathcal{R}}
\newcommand{\allEvents}{\mathcal{E}}
\newcommand{\replicaVar}{r}
\newcommand{\messageVar}{m}
\newcommand{\messageValues}{\mathcal{V}_m}
\newcommand{\messageTypes}{\mathcal{T}_m}
\newcommand{\eventVar}{e}
\newcommand{\eventValues}{\mathcal{V}_e}
\newcommand{\eventTypes}{\mathcal{T}_e}
\newcommand{\sendEventType}{send}
\newcommand{\receiveEventType}{receive}
\newcommand{\internalEventType}{internal}
\newcommand{\historVar}{H}
\newcommand{\messagesPartition}[2]{#1\!\left[#2\right]\!}

% Protocol variables
\newcommand{\protocol}{\mathcal{P}}
\newcommand{\protocolStates}{\Sigma_p}
\newcommand{\protocolInit}{s^0_p}
\newcommand{\protocolTransition}{\delta_p}
\newcommand{\protocolFinal}{F_p}

% Monitor variables
\newcommand{\monitor}{\mu}
\newcommand{\monitorStates}{\Sigma_\mu}
\newcommand{\monitorTransition}{\delta_\mu}
\newcommand{\monitorInit}{s^0_\mu}

% Transition variables
\newcommand{\eventQueue}{E}
\newcommand{\messagePool}{pool}
\newcommand{\statesMap}{states}
\newcommand{\messagesMap}{messages}
\newcommand{\monitorStateVar}{s}

We formalize the semantics of a \toolname unit test for a protocol as a product between a transition system modeling the executions of the protocol with a set $\allReplicas$ of replicas ($\bigl| \allReplicas \bigr| = n$) and a transition system modeling a monitor that controls the delivery of messages to the replicas and asserts some property. The latter is an abstract representation of  \toolname unit tests whose syntax is presented in Section~\ref{syntax}. We characterize the capability of the monitor to restrict the protocol behavior by showing that it can be programmed to reproduce precisely any set of executions that differ only in the order of concurrent computation steps (not related by the standard happens-before relation). 

The set $\allReplicas$ of replicas exchange messages during the execution of the protocol. A message $\messageVar$ is a tuple $(\allReplicas \times \allReplicas \times \messageValues \times \messageTypes)$, where $\messageValues$ denotes the set of possible message values and $\messageTypes$ denotes the set of possible message types. Any set $M \subseteq \allMessages$ of messages can be partitioned based on the replica the message is intended to $\messagesPartition{M}{\replicaVar} = M \cap (\allReplicas \times \left\{\replicaVar\right\} \times \messageValues \times \messageTypes)$ where $r \in \allReplicas$.

An execution of a protocol at a replica can be characterized by a sequence of events. Events can correspond to sending or receiving a message, or an internal computational step. For example, in PBFT an internal step can be committing a value or changing the view. An event $\eventVar$ is a tuple $(\allReplicas \times \eventTypes \times \eventValues)$, where $\eventTypes = \left\{\sendEventType, \receiveEventType, \internalEventType\right\}$ is the set of event types and $\eventValues \supseteq \allMessages$ is the set of possible event values. We say $\allEvents$ is the set of all events.

\subsection{Protocol transition system}
In a protocol, we represent each replica as a transition system $(\protocolStates, \protocolInit, \protocolTransition, \protocolFinal)$ where 
\begin{itemize}
    \item $\protocolStates$ is the set of replica states as defined by the protocol with the initial state $\protocolInit \in \protocolStates$
    \item $\protocolTransition: \protocolStates \times (\allMessages \cup \left\{\bot\right\}) \rightharpoonup \protocolStates \times \allEvents$ is the (partial) transition function of the replica. Each transition emits an event. Internal steps are represented with transitions $\protocolTransition(s, \bot)$ while receiving a message $\messageVar$ is represented with a transition $\protocolTransition(s,m)$. We assume that internal steps and message receive steps cannot be enabled in the same state, i.e. from any state $s \in \protocolStates$ if $\protocolTransition(s, \bot)$ is defined, then $\protocolTransition(s,m)$ for any $m \in \allMessages$ is not defined.
    \item $\protocolFinal \subseteq \protocolStates$. For any state $s \in \protocolFinal$, $\protocolTransition$ is not defined from $s$. Final states allows us to restrict the length of an execution in each replica.
\end{itemize}
For example, in PBFT, the set of states is a valuation of the local variables (index, view, etc) with the initial state as both index and view being 0. A replica in PBFT can transition to a leader state and emit the corresponding event.

A protocol $\protocol$ is a product of $\bigl| \allReplicas \bigr| = n$ replica transitions systems. The configuration of the protocol is denoted by $C = (\eventQueue, \messagePool, \statesMap, \messagesMap)$ where,
\begin{itemize}
    \item $\eventQueue = (e_0, e_1, \cdots)$ is a sequence of events $e_i \in \allEvents$. This serves as a queue of events where replicas push new events to be consumed by the monitor (as it will be clear when defining the monitor transition system below).
    \item $\messagePool \subseteq \allMessages$ is the set of messages in transit between different replicas
    \item $\statesMap$ maps each replica $r$ to a state in $\protocolStates$
    \item $\messagesMap$ maps each replica $r$ to a sequence of messages $(m_0, m_1, \cdots)$ with $m_i \in \messagesPartition{\allMessages}{r}$. This sequence is used as an inbound queue that replicas use to process incoming messages
\end{itemize}

\begin{figure}[h]
\begin{small}
    \centering
    \begin{align*}
        \begin{prooftree}
            \hypo{\protocolTransition(\statesMap\!\left[r\right], \bot) = (s', e)}
            \hypo{e = (r, \internalEventType, v)}
            \infer2[\textsc{Internal}]{(\eventQueue, \messagePool, \statesMap, \messagesMap) \xrightarrow{e} (\eventQueue, \messagePool, \statesMap\!\left[r \rightarrow s'\right], \messagesMap)}
        \end{prooftree} \\ \\
        \begin{prooftree}
            \hypo{\protocolTransition(\statesMap\!\left[r\right], \bot) = (s', e)}
            \hypo{e = (r, \sendEventType, m)}
            \infer2[\textsc{Send}]{(\eventQueue, \messagePool, \statesMap, \messagesMap) \xrightarrow{e} (\eventQueue.e, \messagePool \cup \left\{m\right\}, \statesMap\!\left[r \rightarrow s'\right], \messagesMap)}
        \end{prooftree} \\ \\
        \begin{prooftree}
            \hypo{\messagesMap\!\left[r\right] = \sigma}
            \hypo{m \in \messagesPartition{\messagePool}{r}}
            \infer2[\textsc{Network}]{(\eventQueue, \messagePool, \statesMap, \messagesMap) \xrightarrow{network} (\eventQueue, \messagePool \setminus \left\{m\right\}, \statesMap, \messagesMap\!\left[r \rightarrow \sigma.m\right])}
        \end{prooftree} \\ \\
        \begin{prooftree}
            \hypo{\messagesMap\!\left[r\right] = m.\sigma}
            \hypo{\protocolTransition(\statesMap\!\left[r\right], m) = (s', e)}
            \hypo{e = (r, \receiveEventType, m)}
            \infer3[\textsc{Receive}]{(\eventQueue, \messagePool, \statesMap, \messagesMap) \xrightarrow{e} (\eventQueue.e, \messagePool, \statesMap\!\left[r \rightarrow s'\right], \messagesMap\!\left[r \rightarrow \sigma\right])}
        \end{prooftree} \\ \\
        \begin{prooftree}
            \hypo{M \subseteq \allMessages}
            \infer1[\textsc{Adversary}]{(\eventQueue, \messagePool,\statesMap, \messagesMap) \xrightarrow{adversary} (\eventQueue, M, \statesMap, \messagesMap)}
        \end{prooftree}
    \end{align*}
\end{small}
    \caption{Transition rules of a protocol. For a function $f:A\rightarrow B$, we use $f[a\rightarrow b]$ to denote a function $f':A\rightarrow B$ where $f'(a)=b$ and $f'(a')=f(a')$ for all $a'\neq a$.}
    \label{fig:transition_protocol}
\end{figure}

Figure \ref{fig:transition_protocol} refers to the transition rules between two configurations of the protocol. The rule \textsc{Internal} allows a replica to transition its state and emit \textit{\internalEventType} events. In the \textsc{Send} rule, a replica emits a \textit{\sendEventType} event for a message $m$ and the message is added to $\messagePool$. The \textsc{Network} rule adds messages from the pool to a replica's inbound message queue $\messagesMap\!\left[r\right]$. Rule \textsc{Receive} allows replicas to consume a message from their inbound queues by emitting a \textit{\receiveEventType} event. Additionally, the \textsc{Adversary} rule models adversarial (Byzantine) behavior where the message pool is transformed arbitrarily.

We define an execution $\rho$ as a sequence of transitions between configurations:
$\rho = C_0 \xrightarrow{l_0} C_1 \xrightarrow{l_1} \cdots \xrightarrow{l_{k-1}} C_k$.
The initial configuration is $C_0 = ((), \phi, \statesMap_0, \messagesMap_0)$ with $\forall r \in \allReplicas$, $\statesMap_0\!\left[r\right] = \protocolInit$ and $\messagesMap_0\!\left[r\right] = ()$.
An execution $\rho$ is \emph{complete} if all replicas have reached final states. ($\forall r, \statesMap_k\!\left[r\right] \in \protocolFinal$). 

The \emph{history} of an execution $\rho$ is the tuple $H_{\rho} = (E_{\rho}, <_{\rho})$ where $E_{\rho}$ is the set of events in $\rho$ ordered by the standard (partial) happens-before order $<_{\rho}$. Formally, $E_{\rho} = \left\{e \mid \exists l_i\in \rho, l_i = e\right\}$. We will use $e \in E_{\rho}$ and $e \in H_{\rho}$ interchangeably to denote an event exists in a history. For two events $e_1, e_2 \in E_{\rho}$, we say $e_1 <_{\rho} e_2$ if
(1) $e_1, e_2$ are emitted by the same replica, and $e_1$ occurred before $e_2$ in $\rho$,
(2) $e_1$ is a send event and $e_2$ is the matching receive event, i.e., $e_1 = (m.from, send, m)$ and $e_2 = (m.to, receive, m)$, and 
(2) (transitive closure) there exists $e_3$ such that $e_1 <_{\rho} e_3$ and $e_3 <_{\rho} e_2$.

Also, we define $M_{\rho} = \left\{m \mid (m.to, receive, m) \in E_{\rho}\right\}$ as the set of messages delivered to the replicas in the execution.

\newcommand{\choices}{\mathbb{B}}
\newcommand{\pass}{\bot}
\newcommand{\block}{\top}
\newcommand{\achoice}{b}
\newcommand{\blockedMess}{\mathit{blocked}}

\subsection{Monitor transition system}

\toolname includes a central monitor that receives all communication from the replicas and is able to control the delay of delivered messages or send new messages by itself as a way to model Byzantine faults. The monitor is also used to assert some property.
The central monitor is driven by the events emitted by the replicas. At each step, the monitor can decide to block a certain message from being delivered or to deliver some set of messages based on the current event and context (monitor state). We define the monitor transition system as a tuple $ \monitor = (\monitorStates, \monitorInit, \monitorTransition, F_\mu)$ where
\begin{itemize}
    \item $\monitorStates$ is the set of possible monitor states with $\monitorInit$ as the initial state 
    \item $\monitorTransition: \monitorStates \times \allEvents \rightharpoonup \monitorStates \times 2^{\allMessages}\times \choices$ is the transition function which accepts the current state and event, and transitions to a new state along with a set of messages to be delivered and a Boolean value in $\choices=\{\pass,\block\}$ which in the case of a send event signals whether the message is blocked (if $\block$) or available for delivery (if $\pass$). If the message is blocked in order to be delivered later, then it can be stored in the state of the monitor until delivery. 
    %(we omit a precise formalization since this would be standard). 
    $\monitorTransition$ encapsulates the semantics of executing the filters and state machine for a given unit test.
    \item $F^M \subseteq \monitorStates$ is the set of accepting states (used to signal some property being satisfied)
\end{itemize}
A configuration of the monitor is defined by $C=(\eventQueue, \blockedMess, \messagesMap, \monitorStateVar)$ where
\begin{itemize}
    \item $\eventQueue$ is a sequence of events as described in the protocol transition system. Here it serves the purpose of an event queue for the monitor to consume.
    \item $\blockedMess$ is the set of blocked messages
    \item $\messagesMap$ maps each replica to a sequence of messages as in the protocol transition system
    \item $\monitorStateVar \in \monitorStates$ is a monitor state
\end{itemize}

Figure \ref{fig:transition_monitor} describes the \textsc{Monitor} transition rule. The rule invokes the transition function $\monitorTransition$ with the head of the event queue $\eventQueue$ and the current monitor state as input and
returns the new state, a set of messages to deliver, and a possibly updated set of blocked messages. The delivered messages are added to the respective replica's inbound message queues and the monitor state is updated.

\begin{figure}
\begin{small}
    \centering
    \begin{align*}
        \begin{prooftree}
            \hypo{\eventQueue = e.\eventQueue'\quad\quad \monitorTransition(s, e) = (s', M, \achoice)\quad\quad\forall r.\ \messagesMap'(r) = \messagesMap(r).M[r]}
        \end{prooftree} \\
        \begin{prooftree}
            \hypo{\blockedMess' = \blockedMess \cup ( \achoice?\{m\}:\emptyset), \mbox{ if $e = (\_, \sendEventType, m)$, and $\blockedMess'=\blockedMess$, otherwise} }
            \infer1{(\eventQueue,\blockedMess,\messagesMap,\monitorStateVar) \xrightarrow{monitor} (\eventQueue',\blockedMess',\messagesMap', \monitorStateVar')}
        \end{prooftree}
    \end{align*}
\end{small}
    \caption{Monitor transition rule ( $\mathit{cond}?v_1:v_2$ is interpreted to $v_1$ if $cond$ is true and $v_2$, otherwise).}
    \label{fig:transition_monitor}
\end{figure}

\subsection{Product transition system} 
The asynchronous product of the two transition systems, that of the protocol $\protocol$ and the monitor $\monitor$ defines the set of executions admitted by a unit test (that will be explored by \toolname). The protocol transition system takes steps which publish events that are consumed by steps of the monitor. In turn, the monitor controls the messages that are consumed by the replicas in the protocol transition system. A configuration of the product transition system is $C = (\eventQueue, \messagePool, \statesMap, \messagesMap, \blockedMess, \monitorStateVar)$:
\begin{itemize}
    \item $\eventQueue$, $\messagePool$, $\statesMap$ and $\messagesMap$ are defined as in the protocol transition system.
    \item $\blockedMess$ is the set of blocked messages controlled by the monitor and $s$ is the monitor state.
\end{itemize}

Figure~\ref{fig:transition_product} defines the transition rules for the product transition system. Steps can be either \textsc{Internal, Send, Receive} from the protocol transition system (transitions labeled by events $e \in \allEvents$), a step in the monitor transition system \textsc{Monitor}, or a variation of the \textsc{Network} rule (\textsc{Network-NB}) which adds a \emph{non-blocked} message to a replica's inbound message queue. Note that a monitor step can deliver a set of messages (add them to replica inbound queues) which have been either sent by replicas in the past (these messages were blocked by the monitor and stored in its state) or fictitious messages (corresponding to Byzantine behavior). Therefore, a monitor step simulates a (possibly-empty) sequence of \textsc{Network} and \textsc{Adversary} steps from the protocol transition system. 

\begin{figure}
\begin{small}
    \centering
    \begin{align*}
        \begin{prooftree}
            \hypo{(\eventQueue, \messagePool, \statesMap, \messagesMap) \xrightarrow{e \in \allEvents} (\eventQueue', \messagePool', \statesMap', \messagesMap')}
            \infer1[\textsc{Protocol}]{(\eventQueue,\messagePool, \statesMap, \messagesMap,\blockedMess,\monitorStateVar) \longrightarrow (\eventQueue', \messagePool', \statesMap', \messagesMap', \blockedMess,\monitorStateVar)}
        \end{prooftree} \\ \\
        \begin{prooftree}
            \hypo{(\eventQueue, \blockedMess, \messagesMap, \monitorStateVar) \xrightarrow{monitor} (\eventQueue', \blockedMess', \messagesMap', \monitorStateVar')}
            \infer1[\textsc{Monitor}]{(\eventQueue,\messagePool, \statesMap, \messagesMap,\blockedMess, \monitorStateVar) \longrightarrow (\eventQueue', \messagePool, \statesMap, \messagesMap', \blockedMess', \monitorStateVar')}
        \end{prooftree} \\ \\
        \begin{prooftree}
            \hypo{\messagesMap\!\left[r\right] = \sigma}
            \hypo{m \in \messagesPartition{\messagePool}{r}}
            \hypo{m \not\in \blockedMess}
            \infer3[\textsc{Network-NB}]{\begin{matrix}
                (\eventQueue, \messagePool, \statesMap, \messagesMap,\blockedMess,  \monitorStateVar)\\
                \longrightarrow (\eventQueue, \messagePool \setminus \left\{m\right\}, \statesMap, \messagesMap\!\left[r \rightarrow \sigma.m\right],\blockedMess,\monitorStateVar)
            \end{matrix}
            }
        \end{prooftree}
    \end{align*}
\end{small}
    \caption{Transition rules of product system}
    \label{fig:transition_product}
\end{figure}

A run of the product system $\rho = C_0 \xrightarrow{l_0} C_1 \xrightarrow{l_1} \cdots \xrightarrow{l_{k-1}} C_k$ is a sequence of transitions as above. 
A run is accepting if the monitor's state in the last configuration is a final state, i.e., $C^M_k.s \in F^M$. The set of accepting states of the monitor models the success or failure of a unit test. \toolname uses a randomized exploration algorithm in order to explore the non-determinism introduced by \textsc{Network-NB} transitions.

\subsection{On the Expressivity of the Monitor}

We give a characterization of the monitor's capability to restrict the protocol behavior. We show that as an extreme case, the monitor can restrict the protocol to produce a \emph{single} history for a complete execution, i.e., all the complete executions in the product transition system have the same history. We characterize the capabilities of the monitor in terms of histories because the monitor cannot control the order between concurrent events, which are incomparable w.r.t. happens-before. The order in which such events are pushed to the event queue by the replicas is not under the control of the monitor. In theory, the number of executions that have the same history can still be exponential. However, in practice, for consensus protocols in particular, large numbers of them will be indistinguishable in the sense that every replica will go through exactly the same states (modulo stuttering). For instance, the order between concurrent receive events on different replicas does not affect any replica local state. Since assertions are expressed on local replica states, restrictions in terms of produced histories are effective.

We state our result as a relation between the histories produced in a product transition system and the history of a given complete protocol execution $\rho$. Histories of possibly incomplete executions of the product transition system are not necessarily equal to the history of $\rho$ but only a \emph{prefix}. The prefix relation $\preceq$ between two histories $H_1 = (E_1, <_1)$ and $H_2 = (E_2, <_2)$ is defined as usual, i.e., $H_1 \preceq H_2$ if 
(1) Downward closure: $E_1 \subseteq E_2$ and for every event $e\in E_1$ and $e'\in E_2$, $e' <_2 e \Rightarrow e' \in E_1 \land e' <_1 e$, and (2) Preserving happens before: For two events $e, e' \in E_1$, $e <_1 e' \Leftrightarrow e <_2 e'$.

\begin{theorem}
    \label{thm:main}
    For any complete run $\rho$ in the protocol $\protocol$, there exists a monitor $\monitor$ such that, for all executions $\rho'$ in the product transition system of $\protocol$ and $\monitor$, $H_{\rho'} \preceq H_{\rho}$
\end{theorem}

To prove Theorem~\ref{thm:main}, we construct a monitor to reproduce \emph{exactly one} history. In practice however, a developer writes unit tests to reproduce one of a set of histories. For example, in a unit test of PBFT, a developer is interested in executions where \texttt{Commit} messages are delivered to a replica before \texttt{Prepare} messages, without restricting the order between messages of the same type. Our DSL allows a developer to program such constraints into the monitor's transition function $\monitorTransition$ and hence observe the exact expected behavior. We defer a more elaborate discussion to Section~\ref{syntax}. In the next section we describe the effort needed to instrument an existing protocol implementation with the monitor i.e. \toolname.
\section{\toolname unit tests}
\label{syntax}

\subsection{Instrumentation}
\begin{figure}
  \begin{minipage}{0.5\linewidth}
    \begin{lstlisting}[caption=Initialization code, label=snp:initialization]
server, err := NewStrategyDriver(
  Config{
    APIServerAddr: ":7074",
    NumReplicas: 4,
  },
  PBFTMessageParser(),
  NewPCTStrategyWithTestCase(
    PCTStrategyConfig{
      MaxEvents: 1000, 
      Depth: 10,
    },
    TestCaseOne(),
  ),
  StrategyConfig{
    Iterations: 1000, 
    IterationTimeout: "30s",
  },
)
server.Start()
<-onInterrupt
server.Stop()
    \end{lstlisting}
  \end{minipage}%
  \begin{minipage}{0.02\linewidth}
    \;
  \end{minipage}%
  \begin{minipage}{0.48\linewidth}
    \centering
    \includegraphics[width=\linewidth]{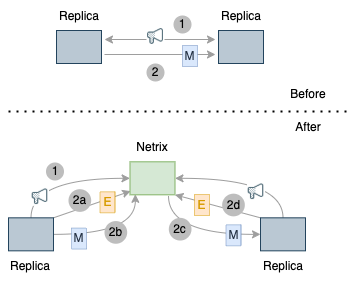}
    \caption{\textbf{Changes to the implementation}}{(1) Discovery process, replicas establish connection to \toolname (\textbf{After}) as opposed to discovering each other (\textbf{Before}). (2) \textbf{Before}: replicas exchange messages. \textbf{After}: (a) Replicas send a message send event to \toolname (b) followed by the message. (c) Replica receives the message and (d) submits a receive event}
    \label{fig:instrumentation}
  \end{minipage}
\end{figure}

Protocol implementations contain communication APIs to facilitate sending/receiving messages. To test the implementation using \toolname, one needs to augment the communication APIs with a Shim. The Shim sends messages to \toolname central server thereby giving it control to decide the order of messages delivered. To aid the instrumentation effort of writing a shim, we provide language specific libraries (currently in \texttt{Go} and \texttt{Java}). Figure~\ref{fig:instrumentation} illustrates the changes needed. The Shim first establishes a connection to \toolname and communicates the relevant information. For every message that the replica sends, the shim should first communicate an event of type message-send tagged with the message identifier and then submit the message to \toolname. Additionally, \toolname can issue directives such as \texttt{start}, \texttt{stop} and \texttt{restart} to each replica. The Shim should communicate the directives to the implementation. Since \toolname is protocol agnostic, each message is recorded as a sequence of bytes. To read the contents of the messages in the unit tests, one needs to provide a serialization and deserialization adapter.

\subsection{Implementation}
The central testing server of \toolname is available as a ''\texttt{go}'' library. The library defines a \texttt{Strategy} interface with a \texttt{Step} function to determine the set of messages to deliver at each step. The interface facilitates write other testing strategies similar to PCTCP. Our approach is implemented in the \texttt{PCTStrategyWithTestCase} strategy. A \texttt{TestCase} object encapsulates a unit test which includes filters and the state machine. \texttt{PCTStrategyWithTestCase} accepts a \texttt{TestCase} as argument. Snippet~\ref{snp:initialization} demonstrates the setup code to run \toolname. As shown, developers should create a strategy driver by invoking \texttt{NewStrategyDriver} and then calling \texttt{driver.Start}. The strategy driver takes as arguments, the strategy (\texttt{PCTStrategyWithTestCase} in our case) and a message parser to interpret the contents of the messages. 

Once started, \toolname waits for replicas to connect. Then, \toolname stores the events received from the replicas in a queue \texttt{EventQueue} and the messages received in a pool \texttt{MessagePool}. Between successive iterations of the strategy, \toolname clears \texttt{EventQueue} and \texttt{MessagePool}, and sends a \texttt{restart} directive to all replicas. Additionally, \toolname creates a \texttt{Context} object that encapsulates a key-value store and contains a reference to the \texttt{MessagePool}. The \texttt{Context} is also reset between successive iterations.

\subsection{Domain specific language}
A unit test in \toolname is represented by a \texttt{TestCase} object. \texttt{NewTestCase} creates a new \texttt{TestCase} object with 3 inputs. First, a timeout that defines the duration of the execution (a bound on the exploration). Second, a state machine that determines if the test is a success or failure. Third, a set of filters that define the unit test. Additionally, one can configure a \texttt{SetupFunc} that contains the test harness (e.g. initial set of client requests). To create a state machine, developers should invoke \texttt{NewStateMachine}. Snippet~\ref{snp:statemachine} defines the state machine from Figure~\ref{fig:pbft_property_sm}. To recall, the state machine encodes a generic safety property. The transitions are labelled with a \textit{condition} and states can be accepting. Similarly, the filters are created using \texttt{NewFilterSet}.  For example, Snippet~\ref{snp:testcase} defines \texttt{TestCaseOne} for PBFT and encodes the filter from Example~\ref{exmp:benign_drop_2}. To recall, the filter drops \texttt{Prepare} messages to replicas $p$,$q$ and $r$. \texttt{filterSet.AddFilter} allows adding more filters. Each filter is of the form \texttt{\textcolor{red}{If}(\textit{condition}).\textcolor{red}{Then}(\textit{actions})}. We define, a \textit{Condition} as a function that accept an event and a context object and returns a boolean and \textit{Action} as a function that accept an event and a context object and return a set of messages. 

\begin{figure}
  \centering
  \includegraphics[width=0.5\linewidth]{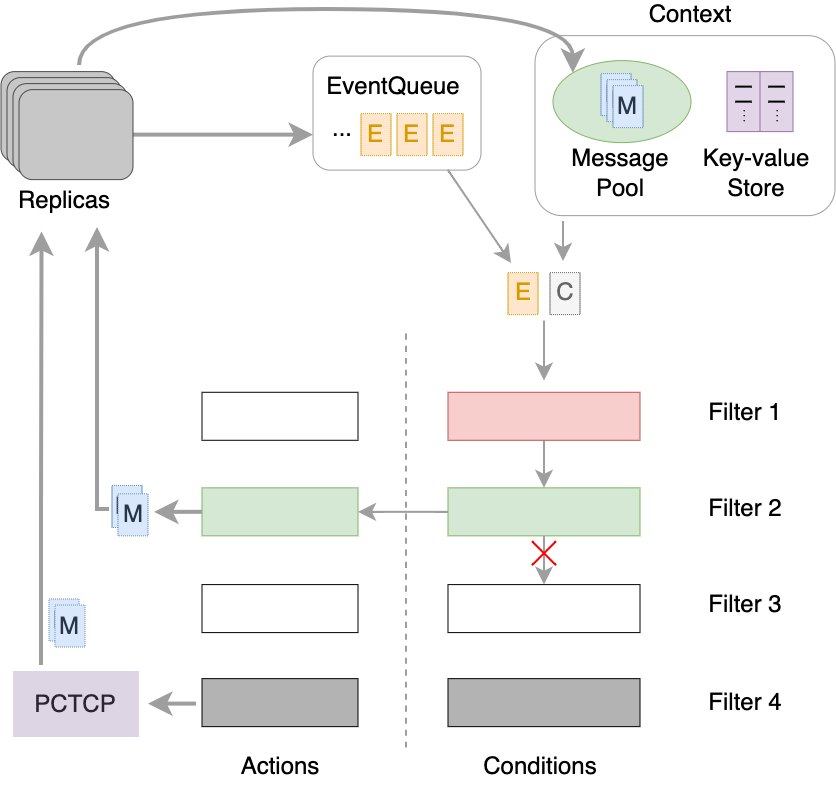}
  \caption{\textbf{Invoking filters with an event and context}: When a filter condition is true for the event E and context C, the messages defined by the action is delivered}
  \label{fig:semantics_filter}
\end{figure}

Internally, we define \texttt{TestCase.Step} which is the concrete implementation of \textsc{Monitor} step defined in Section~\ref{system_model} and represented in Figure~\ref{fig:transition_monitor}. \texttt{TestCase.Step} is invoked for every event in the \texttt{EventQueue} along with a \texttt{Context} object. The \texttt{Context} object is the concrete implementation of a monitor state $s \in \monitorStates$. Similarly, \texttt{Strategy.Step} implements zero or more invocations of the \textsc{Network-NB} transition rule. The number of invocations is determined by the number of messages the underlying randomised algorithm marks for delivery. Figure~\ref{fig:semantics_filter} illustrates the combined semantics. For each event from the \texttt{EventQueue}, we execute the filters similar to a switch case. We check the conditions in the sequential order of the filters and invoke the corresponding actions when a filter condition returns true. When none of the filter conditions match, the event is passed to the random exploration algorithm. Additionally, we deliver messages scheduled by the random exploration algorithm at each step. At the same time, the state machine takes a step for every event. If at the end of the iteration, the state machine is in an accepting state then we consider that iteration of the unit test to be a success and fail otherwise. The state machine is reset to the initial state at the start of every iteration.

\begin{figure}[t]
    \begin{lstlisting}[caption=TestOne, label=snp:testcase,numbers=none,xleftmargin=0in,numbersep=0pt]
function TestOne() *TestCase {
  ...
  filters.AddFilter(
    If( IsMessageType("Prepare").
        And( MessageTo("p").
             Or(MessageTo("q")).
             Or(MessageTo("r")),
      )).Then(DropMessage()) )
  return  NewTestCase("TestOne",sm,filters)
}
      \end{lstlisting}
\end{figure}

  \begin{figure}
    \begin{lstlisting}[caption=State Machine, label=snp:statemachine, language=StateMachine, escapeinside={[}{]},numbers=none,xleftmargin=0in]
sm := NewStateMachine()
initial := sm.Builder()
decidedOnce := initial.On(
  AddToLog([$\alpha$],v,i), 
  "DecidedOnce")
decidedOnce.MarkSuccess()
decidedOnce.On(
  AddToLog([$\beta$],v, i),
  FailState
)
    \end{lstlisting}
  \end{figure}

To reiterate, the syntax for filters is of the general form: \texttt{\textcolor{red}{If}(\textit{condition}).\textcolor{red}{Then}(\textit{actions})}. The core of the DSL defines generic \textit{conditions} (with boolean connectives) and \textit{actions} that are protocol agnostic. We believe, the filters in combination with the \texttt{Context} key-value state allows for describing complex message interleavings. For example, to deliver only a specific number of messages (e.g. 3) of type \texttt{Prepare} to a replica $p$, the developer has to write only two filters. First, to count and deliver if count is less than 3 
\begin{Verbatim}[commandchars=\\\{\}]
  \textcolor{red}{If}(\textcolor{blue}{IsMessageType}("Prepare")
    .\textcolor{blue}{And}(\textcolor{blue}{IsMessageTo}("p).\textcolor{blue}{And}(\textcolor{blue}{Count}("delivered").\textcolor{blue}{Lt}(3)))
    .\textcolor{red}{Then}(\textcolor{violet}{Count}("delivered").\textcolor{violet}{Incr}, \textcolor{violet}{DeliverMessage}))
\end{Verbatim}
and second, to drop when the count is greater than or equal to 3
\begin{Verbatim}[commandchars=\\\{\}]
  \textcolor{red}{If}(\textcolor{blue}{IsMessageType}("Prepare")
    .\textcolor{blue}{And}(\textcolor{blue}{IsMessageTo}("p).\textcolor{blue}{And}(\textcolor{blue}{Count}("delivered").\textcolor{blue}{Gte}(3)))
    .\textcolor{red}{Then}(\textcolor{violet}{DropMessage}))
\end{Verbatim}
Here, there counter is stored in the auxillary key-value store within \texttt{Context}. Similarly, messages can be stored in the context using \texttt{MessageSet} primitives to facilitate reordering. In general, the \textit{conditions} only read from the \texttt{Context} and actions modify the contents of the \texttt{Context}. Table~\ref{tbl:semantics_cond} lists all the \textit{conditions} along with the semantics and Table~\ref{tbl:semantics_actions} lists all the \textit{actions}.

\begin{table}
  \small
  \caption{Semantics of conditions for event $e$ and context $ctx$}
  \label{tbl:semantics_cond}

  \begin{tabular}{lll}
    \toprule
    Condition&Return value\\
    \midrule
    IsEventOf(r)& true if e.replica = r\\
    IsEventType(t)&true if the e.type = t\\
    IsMessageType(t)&true if e.type=send(m)/receive(m) and m.type=t\\ 
    IsMessageSend&true if e.type=send(*)\\
    IsMessageReceive&true if e.type=receive(*)\\
    IsMessageFrom(r)&true if e.type = send(m)/receive(m) and m.from=r\\
    IsMessageTo(r)&true if e.type = send(m)/receive(m) and m.to=r\\
    IsMessageBetween($r_1$, $r_2$)& true if e.type = send(m)/receive(m) and $\left\{m.to, m.from\right\} = \left\{r_1, r_2\right\}$\\
    Count(c).Lt(v)&true if ctx[c] < v\\
    Count(c).Gt(v)&true if ctx[c] > v\\
    Count(c).Leq(v)&true if ctx[c] <= v\\
    Count(c).Gte(v)&true if ctx[c] >= v\\
    MessageSet(s).Contains&true if e.type = send(m)/receive(m) and m $\in$ ctx[s]\\
    c1.And(c2)&true if c1(e,ctx) $\land$ c2(e,ctx)\\
    c1.Or(c2)&true if c1(e,ctx) $\lor$ c2(e,ctx)\\
    c.Not&true if !c(e,ctx)\\
    \bottomrule
  \end{tabular}
\end{table}

\begin{table}
\small
\centering
\caption{Semantics of actions for an event $e$ and context $ctx$}
\label{tbl:semantics_actions}
\begin{tabular}{lll} 
\toprule
Action & Return value & Context changes \\
\midrule
DeliverMessage & \begin{tabular}[c]{@{}l@{}}if e.type=send(m)\\ returns m \\\end{tabular} & - \\
MessageSet(s).Store      & empty set      & \begin{tabular}{@{}l@{}}
      ctx[s] = ctx[s] $\cup$ m \\
      where e.type=send(m)/receive(m)\\
    \end{tabular}\\
DropMessage & empty set & - \\
MessageSet(s).DeliverAll & returns ctx[s] & ctx[s] is set to empty \\
Count(c).Incr  & empty set & ctx[c]++\\
RecordMessageAs(l) & empty set  & ctx[l] = m where e.type = send(m) \\
\bottomrule
\end{tabular}
\end{table}

Apart from the filters, we introduce a builder pattern to create the state machine. Transition labels of the state machine are \textit{conditions}. Invoking \textcolor{red}{sm.Builder} for a state machine \texttt{sm} returns the initial state and transitions are created using \textcolor{brown}{\texttt{On}} function which accepts a condition and the next state as inputs. We refer the reader to Snippet~\ref{snp:statemachine} for an example.

\subsection{Extending the DSL}
\label{sec:dsl_extension}

As they are just functions, it is possible to augment the DSL with protocol specific \textit{conditions} and \textit{actions}. In Section~\ref{our_approach} (Example~\ref{exmp:benign_drop_1} and Figure~\ref{fig:pbft_property_sm}), we refer to PBFT specific conditions such as \texttt{AddToLog}($\alpha$, $v$, $i$) and \texttt{IsMessageOfView}($v$). Such augmentations are necessary to change contents of messages (e.g. \texttt{ChangePrepareToNil} in Example~\ref{exmp:byzantine}). The additional work is justified as it aids in incremental writing of unit tests. 

Apart from the \textit{conditions} and \textit{actions} in Tables~\ref{tbl:semantics_cond} and~\ref{tbl:semantics_actions}, our DSL provides partitions as first class primitives. The goal of partition primitives is to enforce a logical separation between replicas during an execution. Within the \texttt{SetupFunc} of a \texttt{TestCase}, a new partition can be created using \texttt{NewRandomPartition} which accepts a sequence of integers representing the size of each partition. Furthermore, filter conditions can contain \texttt{IsMessageFromPart(p), IsMessageAcrossPartition} and \texttt{IsMessageWithinPartition}. As a special case we define \texttt{IsolateNode(r)} partitions replica $r$ from the rest.
\section{Case studies}
\label{case_study}

Using our DSL we write unit tests for three open source protocol implementations, \textbf{Tendermint}\footnote{https://github.com/tendermint/tendermint}~\cite{tendermint} (version 34.3), \textbf{Raft}\footnote{https://github.com/etcd-io/etcd/tree/main/raft}~\cite{raft} (version v3.5.2) and \textbf{BFTSmart}\footnote{https://github.com/bft-smart/library}. They are production implementations of popular consensus algorithms and are used in a wide variety of applications. The Tendermint protocol is a Byzantine consensus algorithm inspired by PBFT and is the backbone of the cosmos network\footnote{https://cosmos.network}. We test the official implementation of Tendermint written in Go. Similarly, BFTSmart\cite{DBLP:conf/dsn/BessaniSA14,DBLP:conf/edcc/SousaB12} implements in Java a Byzantine consensus algorithm. It is used to build key-value stores and distributed file systems. Raft is a popular benign consensus protocol that tolerates crash failures. We test the Go implementation that is used in many cloud services such as \texttt{etcd} and distributed graph databases such as \texttt{dgraph}. 

The instrumentation effort aided by our language specific libraries needed to test these implementation is very little and is evident in the size of the instrumented code added - 600 LOC for Tendermint codebase of 150kLOC, 120LOC for Raft codebase of 16kLOC and 150LOC for BFTSmart codebase of 16kLOC.

For the three implementations, we write a total of 34 unit tests. In the case of Tendermint (discussed in Section~\ref{eval:tendermint}), we write 15 unit tests and were able to identify 4 new instances where the implementation deviates from the protocol specification. One of the deviations leads to a new liveness bug which increases the time to achieve consensus in the presence of asynchrony by a magnitude of 10 (from ~30secs to ~6mins). The deviations were acknowledged by the Tendermint team and were duly fixed. To strengthen our claim of using \toolname for regression tests, we were able to rerun the same tests without any modification on the updated version and the tests failed to produce a violating execution on the updated version.

To show that \toolname can be used to generically test consensus protocol implementations, we look at Raft (Section~\ref{eval:raft}) and BFTSmart (Section~\ref{eval:bftsmart}). Using an implementation of Raft that aligns with a buggy version of the protocol specification, our unit tests were able to find 4 documented bugs. Note that Raft and Tendermint implementations are both written in Go. We test BFTSmart (written in Java) to show that \toolname can be used to test implementations independent of the programming language. We write 8 unit tests as shown in Table~\ref{tbl:tests}.

To contrast \toolname with a randomized testing algorithms, we ran PCTCP on all three implementations. It fails to capture any of the new bugs/deviations in Tendermint or known bugs in Raft. We believe that the developers intuition is key to catching bugs. However, writing many unit tests requires time and effort/expertise and it is important to balance the tradeoffs. In our case, developer effort translates to number of filters in a unit test. We define a notion of \textit{filter distance} to encode the significance of a filter in a unit test and experimentally show that PCTCP can do the work of shorter distance filters. 

\begin{table}
\small
\centering
\caption{\textbf{List of unit tests}. The table lists unit tests grouped by the protocol. The columns are number of filters, state machine states, LOC and outcomes in number of iterations that successfully caught the interesting scenario/bug. * indicates new bugs found and \^{} indicates tests for replicating known bugs}
\label{tbl:tests}
\resizebox{0.9\linewidth}{!}{%

\begin{tabular}{lcccc|lcccc} 
\toprule
Name                   & \#F & \#S & LOC & Outcomes  & Name              & \#F & \#S & LOC & Outcomes   \\ 
\cline{1-10}
\multicolumn{5}{c|}{\textbf{Tendermint}}           & \multicolumn{5}{c}{\textbf{Raft}}              \\ 
\hdashline[1pt/1pt]
ExpectUnlock*          & 3   & 5   & 90  & 41/100  & Liveness\^{}      & 5   & 3   & 64  & 15/100   \\
Relocked*              & 4   & 5   & 115 & 53/100 & LivenessNoCQ\^{}  & 5   & 3   & 64  & 100/100    \\
LockedCommit*          & 3   & 5   & 85  & 100/100 & NoLiveness\^{}    & 5   & 3   & 33  & 100/100    \\
LaggingReplica*        & 3   & 4   & 71  & 100/100   & ConfChangeBug\^{} & 5   & 2   & 94  & 55/100   \\ 
\cdashline{6-10}[1pt/1pt]
ForeverLaggingReplica* & 5   & 5   & 89  & 100/100   & DropHeartbeat     & 2   & 3   & 69  & 100/100  \\ 
\cdashline{1-5}[1pt/1pt]
RoundSkip              & 3   & 4   & 74  & 90/100  & DropVotes         & 1   & 3   & 44  & 80/100   \\
BlockVotes             & 2   & 3   & 55  & 33/100  & DropFVotes        & 1   & 2   & 57  & 100/100        \\
PrecommitInvariant     & 1   & 3   & 68  & 100/100 & DropAppend        & 1   & 3   & 81  & 100/100  \\
CommitAfterRoundSkip   & 3   & 3   & 82  & 36/100  & ReVote            & 2   & 3   & 54  & 74/100   \\
DifferentDecisions     & 8   & 3   & 180 & 20/100  & ManyReVote        & 2   & 4   & 64  & 92/100   \\
NilPrevotes            & 2   & 3   & 61  & 99/100  & MultiReVote       & 2   & 4   & 60  & 81/100   \\ 
\cline{6-10}
ProposalNilPrevote     & 1   & 3   & 56  & 56/100  & \multicolumn{5}{c}{\textbf{BFTSmart}}          \\ 
\cdashline{6-10}[1pt/1pt]
NotNilDecide           & 2   & 2   & 49  & 100/100 & DPropForP         & 2   & 3   & 60  & 81/100    \\
GarbledMessage         & 1   & 2   & 68  & 30/100  & DPropSame         & 2   & 2   & 40  & 100/100  \\
HigherRound            & 1   & 3   & 91  & 37/100  & DropWrite         & 1   & 2   & 30  & 100/100  \\
                        &     &     &     &         & DropWriteForP     & 1   & 2   & 33  & 89/100   \\
                        &     &     &     &         & ExpectNewEpoch    & 1   & 2   & 28  & 94/100   \\
                        &     &     &     &         & ExpectStop        & 1   & 2   & 38  & 100/100  \\
                        &     &     &     &         & ByzLeaderChange   & 3   & 2   & 46  & 89/100   \\
                        &     &     &     &         & PrevEpochProposal & 2   & 3   & 53  & 99/100   \\
\bottomrule
\end{tabular}
}
\end{table}

\subsection{Filter distance}
\begin{table}
  \centering
  \footnotesize
  \caption{\textbf{List of Filter distances}: unit tests where filter distances are small and PCTCP is able to explore expected re-orderings. Distance -  measures for each filter pair. Outcomes - number of successful iterations when we remove that filter and run PCTCP. We denote distance as an asymptotic measure where $n$ is number of processes and $r$ is the number of rounds.}
  \label{tbl:pctcp_explore}
  \begin{tabular}{lccc}
    \toprule
    Test Name&Protocol&Filter distances&PCTCP outcomes\\
    \midrule
    \multirow{3}{*}{Liveness} & \multirow{3}{*}{Raft}  & $\infty$ & 0 \\
    & & $n^2$ & 0 \\
    & & $n$ & 19/1000 \\ \hline
    \multirow{3}{*}{LivnessNoCQ} & \multirow{3}{*}{Raft} & $\infty$ & 0 \\
    & & $n^2$ & 0 \\
    & & $n$ & 19/1000 \\ \hline
    DropAppend&Raft&$n$&$21/1000$\\  \hline
    ReVote&Raft&$n$&$409/1000$\\  \hline
    ManyReVote&Raft&$2n$&$14/1000$\\  \hline
    MultiReVote&Raft&$3n$&$1/1000$\\  \hline
    RoundSkip&Tendermint&$r\times n$&$6/1000$\\
    \bottomrule
  \end{tabular}
\end{table}

While some of the filters help impose constraints on the execution space that are absolutely necessary (Byzantine faults), the other only aid exploration to obtain better outcomes in terms of number of successful iterations. Allowing PCTCP like algorithms to explore a diverse set of executions is desirable to gain confidence in the correctness of the implementation. We define a distance metric to be associated with filters. The developer can forego writing filters of short distance and expect PCTCP to explore the required message re-orderings.

From our unit tests, we identified three categories the filters can be grouped into --- (1) \textbf{Byzantine} - Filters that introduce Byzantine behavior, (2) \textbf{Drop} - Filters that drop messages, (3) \textbf{Reorder} - Filters that reorder messages. The Reorder filters can be grouped into pairs. Ones that capture the message and store it in a set and ones that release the messages stored in a set. The metric is defined for a pair of capture-release filters and can be determined syntactically. The Drop filters can be considered as a pair, where the release filter releases messages at the end of the execution. In consensus protocols, messages are associated with rounds/phases. We define the distance as the number of messages between the capture and release filter in a normal execution of the protocol (without any faults). For example, if the capture filter corresponds to messages of round $r$ and the release filter corresponds to message of round $r+2$. The distance is then $2n$ where $n$ (linear in number of processes) messages are sent each round. We observe that PCTCP fails to explore executions where the re-orderings are beyond a certain distance threshold. However, the filters with short distance measures (1-2 communication rounds) are not crucial and PCTCP is able to explore the respective re-ordering as shown in Table~\ref{tbl:pctcp_explore}. Due to its inability to introduce Byzantine behavior, PCTCP fails to explore executions which require Byzantine failures. Hence, we define Byzantine filters to have infinite distance.

To recall, PCTCP algorithm observes the set of messages sent during an execution, stores them into a set of (causally-ordered) chains, and delivers messages from these chains according to a random strategy (we give more details later in this section). The probabilistic guarantee of PCTCP exploring executions with $d$-message re-orderings is $\frac{1}{w^2h^{d-1}}$, where $h$ is length of the execution and $w$ the width of the partial order of messages. As evident from the expression, for larger depth $d$, the probability diminishes very quickly. Our notion of filter distance corresponds with the depth of re-ordering. A higher filter distance implies to a larger depth. Therefore, PCTCP fails to explore many executions where the reordering occurs. On the other hand, when the filter distance is small, PCTCP is effective in exploring many executions where the re-ordering occurs.

\subsection{Tendermint}
\label{eval:tendermint}

\subsubsection{Protocol and instrumentation}
Tendermint relies on Gossip protocol~\cite{gossip} for reliable communication. In Tendermint, a set of $n$ validators (with at the most $f < \frac{n}{3}$ faulty) receive requests from clients. Tendermint groups the requests into blocks, and validators should agree on the order of blocks. Each block is associated with a height. In each height, validators exchange messages in rounds. In round $r$ of height $h$, one validator is chosen to be the proposer and sends \texttt{Propose($h,r,bID$)} message. Validators acknowledge the proposal by broadcasting a \texttt{Prevote($h,r,bID$)} message. When a validator $v$ (including the proposer) receives $2f+1$ distinct \texttt{Prevote} messages that match the proposal, the validators lock onto the proposed block $bID$ ($lockedValue_v = bID$) and broadcast a \texttt{Precommit($h,r,bID$)} message. Initially, all validators have $lockedValue_v = nil$. The block is decided (agreed upon) by a validator in height $h$ if in round $r$ it received $2f+1$ distinct \texttt{Precommit} messages that match the proposal.

The existing implementation~\footnote{http://github.com/tendermint/tendermint} of Tendermint is written in \texttt{Go}. We modified the implementation's \texttt{Transport} and \texttt{Connection} interfaces to enable communication with \toolname's API. For theses changes we added 600 LOC to an existing codebase of 150k LOC. Additionally, implementing this was straightforward as the necessary abstractions were already in place.

\subsubsection{Unit tests for Tendermint}

To describe unit test scenarios, we referred to the protocol specification as described in the original paper~\cite{tendermint}, the invariants in the proofs of the protocol, and consulted with the Tendermint developer team.
\paragraph{Interesting scenarios}
\label{tendermint:interesting_scnearios}
The interesting scenarios are motivated by deviations to the synchronous execution path (no failures) of the protocol. For example, when a validator receives 2/3rd \emph{nil} \texttt{Prevote}s, it should send a \emph{nil} \texttt{Precommit}. We simulate this with a filter that drops \texttt{Propose} messages. As a consequence, when a validator does not receive the \texttt{Propose} message, it sends a \emph{nil} \texttt{Prevote}. Additionally, we write a unit test to force the validators to move to round 1 and the state machines asserts that all the validators moved to round 1.

The developer can infer the property to assert, like the scenario, from the protocol specification. It can be generic safety properties (e.g., replicas do not commit on different blocks), or specific properties related to the scenario at hand. For example, if a validator sends a \textit{nil} \texttt{Precommit} upon receiving 2/3rds \emph{nil} \texttt{Prevote}s. We were particularly interested in exploring behaviors where validators moved to higher rounds for two reasons. First, we are able to test for protocol clauses that required messages from more than one round. Second, in production, the validators achieve consensus in the first round due to absence of faults. To explore other scenarios which are not commonly observed in production, we referred to the protocol specification and the invariants in the proof. For example, the protocol defines the following clause:
\begin{verbatim}
  Upon (f+1) messages from a higher round r
    Transition to round r
\end{verbatim}

To simulate this scenario, we isolate one validator $p$ and do not deliver any messages from round 0. We then force the remaining validators to move to round 1. According to the protocol specification, after receiving $f+1$ messages from the round 1, the isolated replica transitions to round 1. For the assertion, we define a custom condition \texttt{ReplicaNewRound(p,r)} which is true when we observe a round change event from the validator $p$ for the round $r$.

\paragraph{\textbf{Liveness bug}} We observe that this unit test fails as the implementation does not behave according to the protocol. The failed scenario demonstrates a liveness bug as the lagging replica fails to catch up to the remaining replicas immediately. The time required for the catchup increases as the gap in rounds between the lagging replica and the remaining replica increases. The Tendermint team has acknowledged the bug and is working on fixing it.

\paragraph{\textbf{Protocol coverage}} We introduce a notion of protocol coverage to justify the expressiveness of the DSL. As mentioned above, a protocol is typically specified as a sequence of ''\texttt{Upon}'' clauses followed by a set of execution rules. A unit test covers a protocol clause if, in the executions explored by the unit test, the clause is satisfied at least in one replica. With our unit tests, we covered all the clauses as defined in the protocol specification for both Tendermint and Raft. 

Apart from the protocol specification, the developer can derive test scenarios from the proofs of the protocols. For example, consider the following inductive invariants used in the proof of the Tendermint protocol\footnote{https://github.com/tendermint/spec/tree/master/ivy-proofs}:
\begin{align*}
  v \neq nil \land &\exists p. precommitted(p,r,v) \rightarrow \\
  &\exists quorum. \forall p. p \in quorum \rightarrow prevoted(p,r,v) \\[-6mm]
\end{align*}
This formula states that if a validator $p$ \texttt{Precommit} a non nil value $v$ in round $r$, then a quorum of validators should have sent \texttt{Prevote} messages for $v$ in round $r$. This is an implication of the form $A \rightarrow B$, and the corresponding unit test contains filters to ensure $\neg B$, i.e., a quorum of validators do not \texttt{Prevote} on the \texttt{Proposed} block, and a state machine that reaches the fail state if it observes $A$, i.e., it observes a \texttt{Precommit} from any validator. 

We write 15 unit tests to describe scenarios that are not commonly observed in production out of which 5 unit tests fail. The failed unit tests demonstrate scenarios where the implementation does not behave as defined by the protocol specification. However, the deviations do not lead to safety violations. As described earlier, among the failed unit tests, one demonstrates a performance bug.

\paragraph{Regression tests}
The Tendermint team independently identified 3 of the deviations captured by our unit tests and were working on changes\footnote{https://github.com/tendermint/tendermint/issues/6849, \newline https://github.com/tendermint/tendermint/issues/6850} to the implementation to correct for them. 3 of the unit tests that failed on an earlier version, succeeded on the fixed newer version of the implementation. We ran the unit tests with no modification on the newer implementation.

\subsubsection{DSL extensions}
\begin{table}
  \centering
  \footnotesize
  \caption{\textbf{List of Tendermint Extensions}: common extensions (conditions and actions) to the DSL for Tendermint.}
  \label{tbl:common_extensions}
  \begin{tabular}{lcp{0.35\linewidth}c}
    \toprule
    Extension&Type&Desc&No of usages\\
    \midrule
    IsMessageFromRound(r) & Condition & true if the message is for a round $r$ & 42 \\
    IsVoteFromPart(part) & Condition & true if the message is prevote or precommit and message is from a partition labelled as part & 27 \\
    IsCommit & Condition & true if the event is committing a block & 16 \\
    IsVoteForProposal(p) & Condition & true if message is a prevote or precommit and is for a particular proposal $p$ & 14 \\
    ChangeVoteToNil & Action & is message is a prevote or precommit, changes the vote value it to nil & 12 \\
    RecordProposal(label) & Action & records the proposal in the context with with the specified label & 11 \\
    \bottomrule
  \end{tabular}
\end{table}

Our corpus of unit tests for Tendermint are compact. Furthermore, they serve as a base repository of common filters that can be used to write more unit tests as the implementation evolves. As mentioned in Section~\ref{sec:dsl_extension}, we extend the DSL to define \textit{conditions} and \textit{actions} specific to Tendermint. Our extensions prove to be useful in writing short unit tests due to their reusability. Table~\ref{tbl:common_extensions} reports some of the common extensions and their usage in filters. 

\subsection{Raft}
\label{eval:raft}

\subsubsection{Protocol and instrumentation}

Raft~\cite{DBLP:conf/usenix/OngaroO14} is a crash fault tolerant benign consensus algorithm. A consensus instance starts with a leader election phase followed by a replication phase. In the leader election phase, a candidate leader requests for votes (\texttt{RequestVote} messages) from all processes. Upon receiving a majority of accepting votes (\texttt{RequestVoteReply}), the candidate transitions to a leader. In the replication phase, a leader receives requests from clients, adds it to the log and replicates the log (\texttt{AppendEntries} message). A request is committed if a majority of processes add it to their respective logs.

We instrument an open source version of Raft (etcd/raft). The Raft implementation exposes the protocol primitives as a library and does not contain any communication primitives. Therefore, to build applications using the Raft implementation we need to introduce a communication layer. To integrate the \toolname library into the communication layer, we added an additional ~120 LOC over the 16k LOC of the implementation.

\subsubsection{Unit tests}

\emph{Interesting scenarios.}
Similar to Tendermint, we systematically explore different fault models to explore deviating behavior. Since the system can tolerate $f$ failures, a developer can write a test to drop $f$ messages. For example, we expect the candidate to transition to a leader despite dropping $f$ \texttt{RequestVoteReply} messages to a candidate. The state machine describing the assertion for this unit test contains two states. From the initial state, we transition to a successful \textit{leader-elected} state when a replica becomes the leader.

\paragraph{Known bugs}
Many safety/liveness bugs have been reported on the raft protocol~\footnote{https://groups.google.com/g/raft-dev/c/t4xj6dJTP6E/m/d2D9LrWRza8J}~\footnote{https://decentralizedthoughts.github.io/2020-12-12-raft-liveness-full-omission/} which lead to liveness and safety bugs. In the liveness bug, the current leader is not able to make progress and the replicas cannot elect a new leader. The safety bug occurs due to concurrent reconfiguration requests and is caused due to a split network with two valid quorums.

The scenario for the liveness bug is as follows. Consider a 5 replica system and replica 1 being the current leader. 1 is disconnected from all other processes except for replica 2 and replica 5 is completely isolated. In this scenario, 2 will not initiate a leader election as it is still connected to the current leader. A new leader (other than 1 or 2) cannot receive a majority vote as 2 is still connected to the current leader. Also, the current leader 1 cannot make progress because it is not connected to a majority of the processes (only 2 is connected). The state machine for this test transitions to success state after observing a threshold of recurrent leader election attempts. The filter drops all messages between disconnected replicas and is activated only when the state machine transitions to a particular state. Note that filters can access the state machine. We believe this allows developers to write more complex filters to explore intricate scenarios.

In the current implementation, two parameters - \texttt{Prevote} and \texttt{CheckQuorum} - when turned on overcome the liveness bug. With our tests, we have been able to check that when the parameters are turned off, the bug is present and when turned on, the bug does not occur.  We skip details of the safety bug for the sake of brevity and refer the reader to the linked bug report. A total of 4 tests describe known bugs.

\subsection{BFTSmart}
\label{eval:bftsmart}
\subsubsection{Protocol and instrumentation} BFT-Smart~\cite{DBLP:conf/dsn/BessaniSA14,DBLP:conf/edcc/SousaB12} is a Byzantine fault tolerant consensus algorithm. The implementation is modular and can be configured to run with either crash fault tolerance or Byzantine fault tolerance. The Byzantine algorithm is analogous to PBFT introduced in~\ref{our_approach}. For the sake of brevity we skip the details of the algorithm.

The BFTSmart implementation\footnote{https://github.com/bft-smart/library} is in Java and has a codebase size of ~16k LOC. Our instrumentation uses the Java client library and consists of ~150LOC changes to the implementation.~\toolname can be used to test implementations written in any language and this is evident from our unit tests for BFTSmart. Additionally, testing BFTSmart helped us develop the Java client library.

\subsubsection{Unit tests} 
The unit tests for BFTSmart are centered around moving processes to a new view. Different tests explore varying approaches of forcing a view change and subsequently check for outcomes by delivering messages from earlier views. For example, delivering a \texttt{PrePrepare} message from an earlier view along with the current one. The assertions (state machines) describe the expected outcomes along with generic safety properties. To simulate such view changes, our filters drop messages, re-order proposals and introduce Byzantine failures.

We do not observe any deviations from the protocol specification of BFTSmart nor did we uncover any bugs. This was expected as BFTSmart is a well tested implementation that is robust and reliable. However, we would like to highlight that the tests allowed us to debug and inspect the implementation and check its conformance with the protocol specification without delving deep into understanding the codebase.

\section{Related work}
\label{related_work}
Testing implementations of distributed systems has received considerable attention over the recent years. Probabilistic Concurrency Testing with Chain Partitioning (PCTCP)~\cite{pctcp,DBLP:journals/pacmpl/OzkanMO19} provides precise probabilistic guarantees about observing every possible order between a fixed number of events. The Jepsen tool~\cite{jepsen} makes it possible to introduce benign faults randomly with a certain frequency, which provides very little control over the outcome of a test. Observing a bug or not depends on the interaction between the fault injection frequency and the scheduler which is not controllable. Jepsen has demonstrated its success empirically by finding bugs in well known systems such as Cassandra~\cite{jepsen_cassandra} and Redis~\cite{jepsen_redis}. Similarly, developers also use simulation-based testing where they run many heavily instrumented replica instances on a single machine and randomly introduce message drops or network partitions. This process can be enhanced with simulated virtual clocks to speed up or slow down replicas~\cite{vrsim}. Both Jepsen and PCTCP does not offer any control over the executions that are explored. Furthermore, unlike in our approach, these techniques cannot explore executions that include byzantine failures.

MoDist~\cite{modist}, SAMC~\cite{samc} and CrystalBall~\cite{DBLP:conf/nsdi/YabandehKKK09} adopt model checking techniques to test distributed systems implementations and exhaustively explore the execution space. These techniques require that tests be run for hours on compute intensive hardware (48 full machine days with SAMC). They cannot deal with Byzantive faults. Moreover, differently from these works, \toolname makes it possible to program the amount of asynchrony or faults in the executions, which simplifies the process of root causing and debugging potential violations.

Our work is inspired by Concurrit~\cite{concurrit}, which enables a similar scenario-based testing approach for \emph{multi-threaded} concurrent programs.
It introduces a DSL that enables developers to define tests where they can control the scheduling between threads with a minimal instrumentation effort. This DSL is specific to multi-threading and very different compared to \toolname's DSL which is specific to testing implementations of consensus protocols. GFuzz~\cite{goConcur} applies the idea of exploring different message orderings between concurrent go channels and has demonstrated success in finding concurrency bugs in actual implementations. P\#~\cite{pSharp} is an actor based programming language that allows developers to write asynchronous systems. P\# is embedded in the C\# programming language and is accompanied by a systematic concurrency testing framework. Similar to GFuzz, P\# explores arbitrary event orderings between the actors to find concurrency bugs. However, both GFuzz and P\# do not allow describing specific scenarios to test.

Our DSL primitives are motivated by specification languages for protocols. DISTAL~\cite{distal} programs are a sequence of \textit{Upon} clauses. Each \textit{Upon} is followed by a predicate on the state of the protocol and current message. Similar to our DSL, DISTAL predicates contain counting, sets of messages and comparing message types. ModP~\cite{modP} language allows protocol designers to describe and test a model of the protocol. Similar to DISTAL, ModP machines contains a sequence of \textit{on} event handlers that modify the state of the machine. The \textit{on} handlers are followed by predicates similar to DISTAL. ModP also generates code for testing the programs. While these are effective in finding bugs in a \emph{model} of a protocol, the results however do not help in testing production implementations. The main reason they do not help in production environments is because  model checkers do not scale when applied directly on implementation of large systems. Moreover most model-checkers that work at the programming language level, are not even applicable to this application domain,  as they  focus on primitives for shared memory and not message passing~\cite{rust-shuttle, kotlin-lincheck}. Therefore our DSL provides the only guided way to do exploration of the execution space on implementations.  

\paragraph{Comparison in instrumentation effort}
MoDist adds an interposition layer between the replica and the operating system and it is limited to distributed systems that run on the Windows operating system. Similarly, Jepsen relies on manipulating \texttt{iptables} rules to control communication between replicas. Jepsen, PCTCP, and MoDist require developers to write initialization scripts. Replicas are instantiated using the scripts and allow the tools to control (1) replica processes and (2) the communication between replicas. \toolname like PCTCP requires complete control over the messages. Our network infrastructure allows replicas to communicate the messages via RPC and it is a general infrastructure for any exploration tool that requires control over the communication network.
\section{Conclusion}
\label{conclusion} 
Distributed systems suffer from complex and intricate bugs. Existing testing tools for distributed systems focus on either exhaustive enumeration techniques or using controlled randomness. Therefore bugs are not reproducible and furthermore, they fail to incorporate the semantics of the protocol. With \toolname we aim to raise the level of abstraction in unit tests. We combine programmer provided semantic insights with controlled randomized exploration. This allows tests to be effective by guiding the automatic search towards interesting sensations, while maintaining reproducibility. 

We build an open source tool \toolname that is based on the domain specific language. We use \toolname to test Tendermint~\cite{tendermint}, Raft~\cite{raft} and BFTSmart~\cite{DBLP:conf/dsn/BessaniSA14,DBLP:conf/edcc/SousaB12}. Our unit tests are effective in capturing deviations from the specification in the implementations. Furthermore, we demonstrate the re-usability of the tests by running them on different versions of the implementations and checking the bug fixes made in the implementation.

\bibliographystyle{plainurl}
\bibliography{paper, dblp}
\appendix

\end{document}